\documentclass[10pt, conference, letterpaper]{IEEEtran}
\IEEEoverridecommandlockouts
\usepackage{cite}
\usepackage{amsmath,amssymb,amsfonts}
\usepackage{algorithmic}
\usepackage{graphicx}
\usepackage{textcomp}
\usepackage{bm}
\usepackage{color}
\usepackage{url}
\allowdisplaybreaks

\newtheorem{lemma}{Lemma}
\newtheorem{theorem}{Theorem}

\newtheorem{fact}{Fact}

\newcommand\tab[1][1cm]{\hspace*{#1}}

\newcommand{\braces}[1]{\left\lbrace #1 \right\rbrace}

\def\BibTeX{{\rm B\kern-.05em{\sc i\kern-.025em b}\kern-.08em
    T\kern-.1667em\lower.7ex\hbox{E}\kern-.125emX}}
\begin{document}

\title{Small-Scale Markets for Bilateral Resource Trading in the
  Sharing Economy\\
\thanks{Research was funded in part by NSF grants CNS-1149458, AST 1443891, EFRI-1440969, AST-1343381, AST-1516075, IIS-1538827 and EPCN-1608361. Any opinions, findings, and conclusions or recommendations expressed in this material are those of the authors and do not necessarily reflect the views of the funding agencies.}
}
\author{
 \IEEEauthorblockN{Bainan Xia}
 \IEEEauthorblockA{\textit{Dept. of ECE } \\
 \textit{Texas A\&M University}\\
 Email: xiabainan@tamu.edu}
 \and
 \IEEEauthorblockN{Srinivas Shakkottai}
 \IEEEauthorblockA{\textit{Dept. of ECE } \\
 \textit{Texas A\&M University}\\
 Email: sshakkot@tamu.edu}
 \and
 \IEEEauthorblockN{Vijay Subramanian}
 \IEEEauthorblockA{\textit{Dept. of EECS} \\
 \textit{University of Michigan}\\
 Email: vgsubram@umich.edu}
}
\maketitle
\begin{abstract}
We consider a general small-scale market for agent-to-agent resource
sharing, in which each agent could either be a server (seller) or a
client (buyer) in each time period. In every time period, a server has
a certain amount of resources that any client could consume, and
randomly gets matched with a client. Our target is to maximize the
resource utilization in such an agent-to-agent market, where the
agents are strategic. During each transaction, the server gets money
and the client gets resources. Hence, trade ratio maximization implies
efficiency maximization of our system. We model the proposed market
system through a Mean Field Game approach and prove the existence of
the Mean Field Equilibrium, which can achieve an almost 100\% trade
ratio. Finally, we carry out a simulation study motivated by an agent-to-agent computing market, and a case study on a proposed photovoltaic market, and show the designed market benefits both individuals and the system as a whole.
\end{abstract}


\section{Introduction}\label{sec:intro}
The sharing economy is a paradigm shift in the working of the twenty-first century marketplace.  Supported by the ease of communication and availability of information provided by the Internet, this marketplace innovation has blurred the line between producers and consumers, turning participants into \emph{prosumers} who can both provide and utilize resources and services.   Successful platforms in this space enable access to resources that are commonplace, but are needed at the right place at the right time.  Typically, these resources have the property that ``unused value is wasted value,'' in that idle time cannot be utilized later on.   Examples include P2P networks such as BitTorrent (bartering of bandwidth), Fon (token-based WiFi sharing), Uber/Lyft (typically, fixed-price car sharing), and Airbnb (marketplace-mediated home sharing).

Prosumers typically provide or consume small amounts of resources, which means that bilateral trade (one-to-one) is the norm.  Thus, the sharing platform enables bilateral trading, with options ranging from barter to bargaining with monetary instruments.  Prosumers are ephemeral in that they might participate for some duration of time, and then switch to some other platfrom or stop altogether.  The number of participants at any time is large, which is how sharing systems manage to match demand and supply.  Also, in most existing sharing systems, prosumers act largely as consumers or producers, but rarely switch roles. 
%

A novel set of applications are now emerging in which prosumers switch roles from being producers to consumers frequently.  Here, agents have either demand or resources that are bursty, which results in recurring role changes. Like P2P networks used for content sharing, these applications are associated with easily sharable resources, and provide services that are indistinguishable from traditional sources.  We consider \emph{agent-to-agent} (A2A) market design with two such applications in mind, namely, (i) sharing computational resources and (ii) sharing electricity resources, with details as follows:\\
(a) \emph{Distributed Computation:}  We consider enterprise level systems, such as compute clusters at universities and other data analytics organizations.  Many run with bursty utilization (our data indicates that several clusters at the authors' organization are utilized at about 50-60\% on average), with users often needing more or less resources than available at a single cluster.  Virtualization for sharing is straightforward, and the altenative is to pay a fee for commercial cloud computing services from an organization such as Amazon EC2.\\
(b) \emph{Distributed Electricity Generation:}  We consider rooftop-photovoltaics (PV) based electricity generation at the level of homes and small businesses, where generation depends on the intensity of sunshine.  As we will see, geographic vagaries mean that one can shift from being a producer to consumer often, and the existing grid can allow incorporation of these resources.  Again, a traditional alternative exists in the form of electricity purchase from a utility provider.

While there are existing platforms using a two-sided market approach for some applications, approaches that focus on prosumers that frequently change roles from provider to consumer are few.  Consider a bilateral market in which currency is used as the instrument of trading.  A simple mechanism is one which a consumer (that we term as a \emph{client}) is matched to a random producer (that we term as a \emph{server}), each places a bid, and a trade happens if the client bids higher than the server's demand.  The server then receives the currency equal to her bid, and must incur a cost of providing service.  Thus, an agent in a client role pays the agent in a server role to obtain resources.  Likewise, the agent that is currently a server can use these currency units to obtain resources when it in turn becomes a client.  Also, each time an agent in a client role obtains resources, it generates surplus, measured in currency units.   This corresponds, for example, to the productivity gains due to obtaining compute cycles or electricity.   If the client does not succeed in obtaining service under the sharing economy, it faces an cost, which can be thought of as the negative feeling of having to search for an alternative source or to experience delays.  Would such a market be sustainable, i.e., would there be enough resource trades generating currency (surplus) such that available resource utilization is high? 
 
In this paper, we develop a game theoretic framework to model and analyze A2A markets under the mechanism described above.  The choice of mechanism in a successful marketplace often depends on the timescale of resource usage, with simple solutions such as bartering being effective at fast timescales, and more complex ones like bargaining at long timescales.  The timescale of our candidate applications (minutes to hours) suggest that a low complexity solution is desirable, and the value of the mechanism will be apparent in later sections.  In the context of our applications, random matching of agents is viable since cloud computing is essentially agnostic to geography, while integration of renewable energy into the electricity grid is already well established in the US (eg. using net-metering in which customers can sell back excess renewable energy generated \cite{darghouth2011impact}).  

Our market model consists of random matching between a large number of ephemeral agents that might leave at any time.  We assume that a departing agent is replaced with a new agent, keeping the total number of agents fixed.  The state of any agent is the amount of currency that it possesses at that time.  A client can be constrained to only place a bid if it has sufficient currency to do so.  It is clear that such a budget constraint might restrict entering agents from obtaining resources, and result in low trade volume.  Indeed, some P2P networks such as BitTorrent build in a measure of altruism to reduce friction in the system.  In our context, we also consider models in which the client can obtain a loan from a central entity (a bank loan) or from the server itself (a peer loan).  The client must pay back the loan with interest after the trade using the currency (surplus) generated by receiving resources. 

\subsection{Mean Field Games}

We investigate the existence of an equilibrium using the framework of Mean Field Games (MFG) \cite{LasLio07}.  Here, each agent assumes that the matched agent would play an action drawn \emph{independently} from a fixed distribution over its bid space.  The agent then chooses an action that is the best response against actions drawn in this manner.  The system is said to be at Mean Field Equilibrium (MFE) if this best response action is itself a sample drawn from the assumed bid distribution.  The framework considerably reduces computational overhead, and can easily be shown to be an accurate approximation when the number of agents is asymptotically large in  arnge of applications \cite{IyeJoh14,GumKey13,LiRaj16} including our context.  


The MFG framework offers a relatively simple way of modeling and analyzing large scale games when each subset of agents interacts infrequently.  In the context of the sharing economy, a  particular producer and consumer would rarely be matched together multiple times in their lifetimes, since the number of participants is large and participant lifetime is limited.  This implies that little utility is lost due to minimal history retention, and action choice becomes less complex. 

The main related papers in the MFG setting are \cite{IyeJoh14,GumKey13}.   In\cite{IyeJoh14}, a sytem for auctioning advertisements on a webpage is considered.  Here agents are advertisers that bid for these spots, and the main result shows how convergence to the MFE takes place while learning about the value of winning a slot on the webpage.   The model is extended in \cite{GumKey13} to include hard budget constraints in the sense that agents may only bid an amount less than their existing budget.  The budget itself is updated according to an independent arrival process, and the result is a characterization of the reduced bid that would be made in this case.   Neither of these considers matching markets of producers and consumers that are interchangeable.

\subsection{Other Related Work}

There has recently been much work in the context of the sharing economy, but little in the way of understanding systems in which agents change their roles often.  Most work that deals with this problem considers the special case of data/spectrum sharing in wireless networks.  For instance, \cite{AfrGue12} study pricing models for a system like Fon in which WiFi is shared.   In the same manner, \cite{GaoWan11,GaoIos14} study spectrum sharing and mobile data offload in which peers can use each other's resources in the setting of a small number of agents.  They consider mechanisms across a small number of agents such as contracts and double auctions.  However, they do not consider repeated play with learning of behaviors.  

In the context of electricity resources, \cite{KalWu16}  consider sharing of storage resources in a smartgrid setting, with charging when prices are low and sharing when proces are high.  Unlike their setting, we do not assume any storage, and usage by a customer can only happen with successful  trade.

To the best of our knowledge, there is no prior work that considers mechanism design for bilateral (A2A) repeated games with role switching agents between producer (server) and consumer (client) in the mean field setting.








\subsection{Main Results}


Our main result is a characterization of the mean field equilibrium bid distribution under three systems, namely (i) a hard budget constraint, (ii) a bank loan, and (iii) a peer loan.  We show the existence of a mean field equilibrium in each case, and show that there exists a set of equilibria that are simple, and characterized by the server setting a fixed price $k,$ while the client chooses whether or not to bid $k$ based on her budget and estimate of future value.  In particular, the client decision turns out to be a set of divisions of the budget into intervals, with $k$ being optimal in some and $0$ being optimal in others.  In all cases, if the budget is sufficiently large, the client always bids $k,$ while if it is sufficiently small, it always bids $0.$ 

The stable bid $k$ is not unique, and a set of such bids exist (with the minimum being lower bounded by server cost, and the maximum being upper bounded by the client surplus plus cost of not obtaining service), each one of which is a MFE.  However, the fraction of time that a trade happens (i.e, the client actually bids $k$) is not the same for all systems and values of $k.$  In particular, the bank loan and peer loan models both attain higher trade ratios, particularly in the case when the initial budget of an agent is low.  Essentially, a small boost in the form of a loan (which is retuned immediately with interest via the surplus generated by the trade) is successful in reducing friction in the market allowing it to attain high efficiency.  

The trade ratio also depends on the value of $k$ itself.  Interestingly, maximum trade is not necessarily attained at the lowest possible value of $k,$ but there exists a value between the highest and lowest at which this happens.  The reason is that since clients and servers are interchangeable, extraction of surplus by a server is not always a bad thing from the client's perspective, since it too will gain when it in turn becomes a server.  When the initial budget is low, a client is forced to take a loan in order to bid a high value of $k.$  But when it does so, it transfers a larger sum to the server, which then (subtracting service cost), might be in a position to obtain service without having to take a loan in the future (when it's a client).  Thus, aggregation of surplus at servers may not be bad.


We conduct numerical studies to illustrate the viability of our scheme.  We first run simulations modeling a A2A cloud computing application, in which agents alternate probabilistically between a client and a server.   Insights on  trade ratios and optimal prices are observed in this setting.  We then conduct a case study on the PV electricity generating and trading market.  Here, we use weather, price and demand data of two cities, Austin and El Paso, Texas USA, to show how the system might perform in a realistic setting, and estimate the gains that could be obtained on a per agent (household or small business) basis amount to several hundred dollars a year.
\section{Mean Field Model}\label{sec:model}
We consider a general model of the proposed market with a large number of agents.  Each agent maintains a private budget state and can bid any value within her budget in the role of a client.  When a client gets matched to a server, each places a bid.  If the server indicates a  lower price than what the client proposes, a bilateral trade happens.  At the end of a successful trade, the client pays the server's asking price, receives service and translates it into a dollar value surplus that directly increases her budget.  Meanwhile, the server receives the payment, and pays the cost of the providing service.   Thus, the client will bid strategically under some belief about the likely bids of the server, and \emph{vice versa}.

\begin{figure}[htbp]
\centering
\includegraphics[width=3.5in]{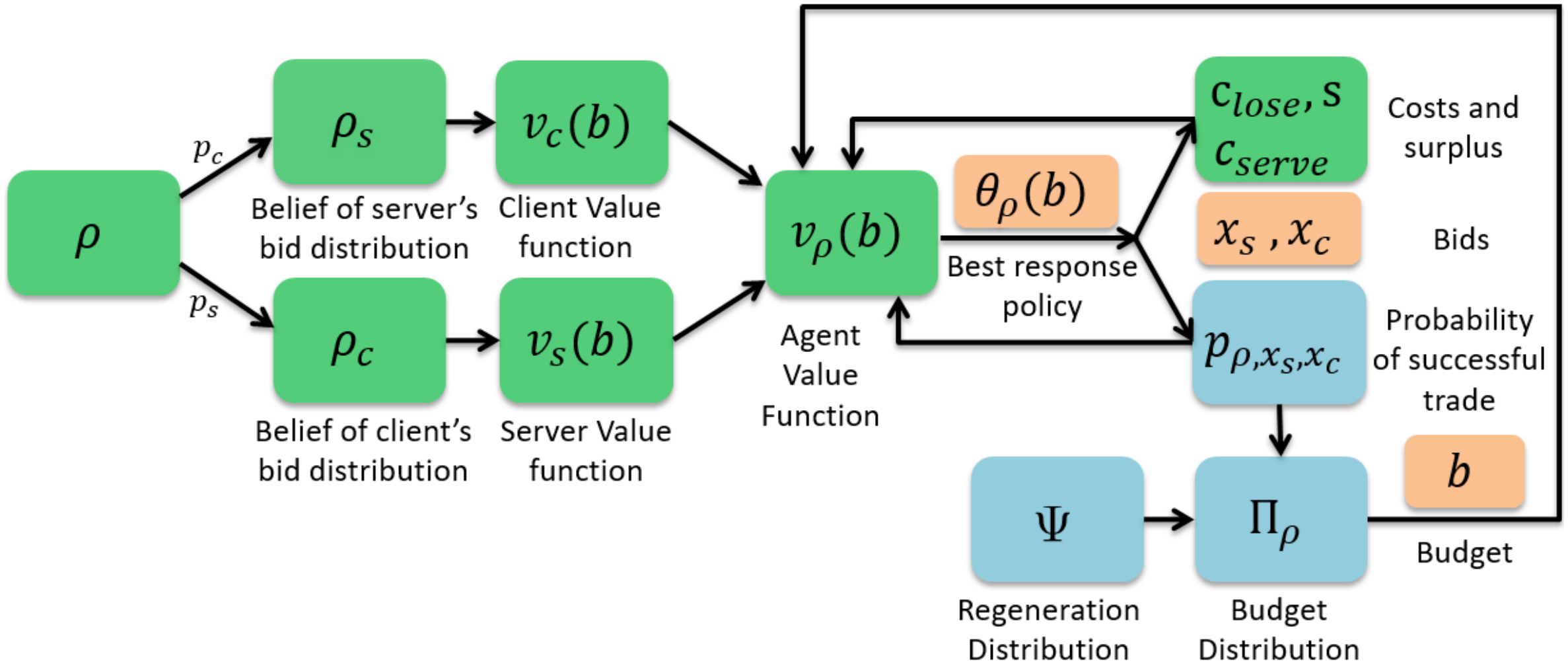}
\caption{Mean Field Game}
\label{fig:MFE_Block}
\vspace{-0.3in}
\end{figure}

Computing the perfect Bayesian equilibrium of such a system is complex, particularly when the number of agents is large.  Instead, we use a mean field approximation of the proposed market system, which has proven to be an accurate representation when the number of agents is asymptotically large~\cite{IyeJoh14,GraMel94}. Fig.~\ref{fig:MFE_Block} illustrates our mean field model from the perspective of a single agent.

At each discrete time step, an agent could either be a client or a server with fixed probabilities, $p_c$ and $p_s$, respectively.  A client places a bid based on her belief of server's bid distribution, and \emph{vice versa}.  Since the number of the agents is large, which implies that both the number of clients and servers at any instant are large as well, each individual can assume that her opponent's bid is drawn independently from the c.d.f. $\rho_{c}$ or $\rho_{s}$, respectively. Then the complexity of the single agent decision making problem is much reduced. In the rest of this section, we will provide a term-wise description of our mean field model with the accompanying notation.

\textbf{Time:} Time is discrete and indexed by $t\in\{0,1,...\}$

\textbf{Agent:} At each time period $t$, an agent could either be a client or a server, with probability $p_c$ or $p_s,$ respectively. Note that the total number of agents is large and $p_c+p_s=1$.

\textbf{Bids:} When a client is matched with a server, each places a bid, denoted as $x_c$ and $x_s,$ for the client and server, respectively.  When $x_c\geq x_s$, the trade succeeds and the client pays $x_s$ to the server.

\textbf{State:} Each agent keeps track of her budget, $b$, as a private state.   At time $t$, the budget of a agent is updated as following if a trade happens, i.e. $x_c\geq x_s$,
\begin{equation}
\label{eq:state_trans}
b[t+1]=
\begin{cases}
\begin{aligned}
&b[t]+s-x_s-\alpha(x_s-b[t])^+, & \text{as a client w.p. }p_c \\
&b[t]-c_{serve}+x_s, & \text{as a server w.p. }p_s \\
\end{aligned}
\end{cases}
\end{equation}
if a trade fails, i.e. $x_c<x_s, b[t+1]=b[t]$\\
Note that $s$ represents the fixed dollar value surplus that a client gains from receiving service, and $c_{serve}$ is the corresponding fixed cost that a server pays for providing service.  Parameter $\alpha$ is the penalty term when a client overdraws on her budget.  One should think of this as taking a loan from a bank (which is our focus case; a peer loan is also possible), which is then paid back in full with interest after service is obtained and surplus is generated.  Hence, $\alpha$ is typically greater than $1$.  The support of the budget is $\mathbb{R}_+:=[0,+\infty)$.

\textbf{Costs:} We have already mentioned $c_{serve},$ which is the server's cost of providing service.  In addition, we introduce another cost, $c_{lose},$ which denotes the cost of failure to obtain service as a client.  This models an instantaneous dissatisfaction suffered by the client, but does not impact her budget.

\textbf{Regeneration:} An agent may quit the system at the beginning of any time period $t$ with probability $1-\beta,$ and may stay with probability $\beta\in(0,1).$ We assume that a new agent enters the system when an old agent leaves, and the budget of the new agent is drawn from a probability distribution $\Psi$  with a density on $B_{init}$ that is a bounded subset of $\mathbb{R}_+$.

\textbf{Best Response Policy:} As an agent participates in the system, she places bids at each time period.  Hence, she needs to solve a repeated decision making problem, given the private budget $b$ and the public belief about the bid distribution $\rho=[\rho_c,\rho_s]$.  The probability of the trade happening can be computed for a given bid using the public belief about her opponent's bid distribution.  This probability characterizes the next step transition of an agent's budget.  Hence, a dynamic program is defined for an agent to find her best response policy, $\theta_{\rho},$ as is shown in green/dark blocks of Fig.~\ref{fig:MFE_Block}. We will discuss this in detail in section \ref{sec:br}.

\textbf{Stationary Distribution of Budget:} Given the best response policy, the state transition of an agent is described by equation \eqref{eq:state_trans} together with the regeneration, which forms the transition kernel of a discrete-time Markov Chain. The stationary distribution of this Markov Chain, $\pi_{\rho},$ is equivalent to the resulting budget distribution after a one-step transition over infinite number of agents with the public belief $\rho$.

\textbf{Mean Field Equilibrium:} Given the assumed bid distribution $\rho$, solving the dynamic program, the best response policy is obtained, which defines the kernel of the budget Markov Chain. Thereafter, taking the stationary distribution of budget together with the best response of each state, a new bid distribution $\gamma(\rho)$ can be calculated. If $\gamma(\rho)$ turns out to be the same as the public belief $\rho$, the system is at an MFE.  Detailed discussions of MFE can be found in Section~\ref{sec:mfe}.

\section{Best Response Policy}\label{sec:br}
To characterize the best response policy of an agent in different roles, we need to formulate the dynamic program mentioned in the previous section. Before that, we first introduce a few easily established facts regarding equilibrium behavior of the agents, which will shrink the space of dynamic programs of interests. Here, we consider the agents bid discrete values in $\mathbb{R}_+$. Later, we will show a specific class of equilibria exists and achieves high efficiency, in which all servers bid one and the same price. We further define random variables $\tilde{X}_s$ and $\tilde{X}_c$ distributed according to $\rho_s$ and $\rho_c$, and the corresponding p.m.f. are $p_{\tilde{X}_s}$ and $p_{\tilde{X}_c}$. Suppose the bid spaces of clients and servers are upper bounded by $\bar{x}_c$ and $\bar{x}_s$, we have following:
\begin{fact} \label{fact1}
Given $\rho_s$, a client should never bid higher than $\bar{x}_s$, where $\bar{x}_s$ is the upper-end of the support of $\rho_s$, i.e., $\rho_s(x_s)=1\ \forall x_s\geq\bar{x}_s$ and $\rho_s(x_s)<1\ \forall x_s<\bar{x}_s$.
\end{fact}

\begin{fact} \label{fact2}
Given $\rho_c$, a server should never bid higher than $\bar{x}_c$, where $\bar{x}_c$ is the upper-end of the support of $\rho_c$, i.e., $\rho_c(x_c)=1\ \forall x_c\geq\bar{x}_c$ and $\rho_c(x_c)<1\ \forall x_c<\bar{x}_c$.
\end{fact}

\begin{fact} \label{fact3}
Given $p_{\tilde{X}_s}$, a client should never bid $x_c$ for $x_c>0$ such that $p_{\tilde{X}_s}(x_c)=0$.
\end{fact}

\begin{fact} \label{fact4}
Given $p_{\tilde{X}_c}$, a server should never bid $x_s$ such that $p_{\tilde{X}_c}(x_s)=0$.
\end{fact}

These facts hold since the violation each of them yields a non-positive expected payoff to a generic agent. Thus, we claim that if an equilibrium exists, which we will discuss in section \ref{sec:mfe}, then in each equilibrium, by Fact \ref{fact1} and \ref{fact2}, we have $\bar{x}_s=\bar{x}_c$, meanwhile, by Fact \ref{fact3} and \ref{fact4}, we have the action space of an agent in each role has the same discrete support, denoted as $\mathcal{D}\subset\mathbb{R}_+$, with the corresponding beliefs.

\subsection{Value Function}
As we discussed in section \ref{sec:model}, the repeated decision making problem for a single agent (with a geometrically distributed lifetime) forms a discounted cost dynamic program. Unlike a traditional market, the agent plays two roles (client or server) probabilistically over its lifetime.  In each role, the agent encounters a different decision making problem, but based on its private budget that is common to both roles.  We track a generic agent just before her role (client or server) is revealed, and we will use this sampling point in the rest of the document.  However, as an agent takes an action (bids) only after her role is revealed, this is consistent with our set-up mentioned in Section~\ref{sec:model}.  Hence, we define the Bellman equation associated with the dynamic program of interest as follows:
\begin{align}
&v_{\rho}(b) =p_s v_s(b) + p_c v_c(b) \nonumber\\
&=p_s \big ( \max_{x_s\in\mathcal{D}}\mathbb{E}_{\rho}[\bm{1}_{\tilde{x}_c \geq x_s}(\beta v_{\rho}(b+x_s-c_{serve})\nonumber\\
&+x_s-c_{serve})+\bm{1}_{\tilde{x}_c < x_s}\beta v_{\rho}(b)] \big) \nonumber\\
&+ p_c \big ( \max_{x_c\in\mathcal{D}\cap[0,b+s/(1+\alpha)]}\mathbb{E}_{\rho}[\bm{1}_{x_c \geq \tilde{x}_s}(\beta v_{\rho}(b+s-\tilde{x}_s\nonumber\\
&-\alpha(\tilde{x}_s-b)^+)+s-\tilde{x}_s)+\bm{1}_{x_c < \tilde{x}_s}(\beta v_{\rho}(b)-c_{lose})] \big), 
\label{eq:value_function}
\end{align}
where $\tilde{x}_s$ and $\tilde{x}_c$ are realizations of random variables, $\tilde{X}_s$ and $\tilde{X}_c$, and $v_\rho(\cdot)$ is the value function of a generic agent, which is in turn composed of an average of $v_s(\cdot)$ and $v_c(\cdot)$ (value functions once her role is revealed).  Since each agent's role is determined exogenously at the beginning of a time period, the evolution of both $v_s(\cdot)$ and $v_c(\cdot)$ depends on $v_\rho(\cdot)$ so that \eqref{eq:value_function} remains consistent. We believe that the exogenously driven role choice makes this a natural assumption.  Since the role of the agent in our context is usually determined by the external environment, the value of currency should be determined by the underlying market and not the role one's currently playing.

In our model, the client is allowed to overdraw her budget with an upper limit such that the budget does not end up negative after any possible transaction, i.e. the client is allowed to choose up to the maximum value of $x_c$ subject to  $(b+s-x_c-\alpha(x_c-b)^+)$ being non-negative. A simple calculation then yields the upper limit of a client's bid as $b+s/(1+\alpha)$. Also, for both clients and servers, when a trade happens, a budget update as well as an instantaneous gain in value is induced, which captures the fact that the trade generates value both in the present and in the future.  Further, notice that the expectation of the indicator functions in equation (\ref{eq:value_function}) can be determined using the probability of trade happening, which in turn can be calculated directly using $\rho$. Then we can further characterize $v_{\rho}(b)$ as follows:
\begin{align}
&v_{\rho}(b)=\nonumber\\
&= p_s \Big ( \beta v_{\rho}(b)+\max_{x_s\in\mathcal{D}}(1-\rho_c(x_s))(\beta(v_{\rho}(b+x_s-c_{serve})\nonumber\\
&-v_{\rho}(b))+x_s-c_{serve}) \Big) + p_c \Big( \max_{x_c\in\mathcal{D}\cap[0,b+s/(1+\alpha)]}\Big[ \nonumber\\
&\sum_{\tilde{x}_s=0}^{x_c}p_{\tilde{X}_s}(\tilde{x}_s)\big(\beta v_{\rho}(b+s-\tilde{x}_s-\alpha(\tilde{x}_s-b)^+)+s-\tilde{x}_s\big) \nonumber\\
&+(1-\rho_s(x_c))(\beta v_{\rho}(b)-c_{lose}) \Big ] \Big) \nonumber \displaybreak[0] \\
&= \beta v_{\rho}(b) + \max_{(x_s,x_c)\in\mathcal{A}(b)} \Big( \Big[  p_s(1-\rho_c(x_s))(x_s-c_{serve}) \nonumber \\
&+ p_c \big(\rho_s(x_c)(s-\mathbb{E}[\tilde{X}_s|\tilde{X}_s\leq x_c])- (1-\rho_s(x_c))c_{lose} \big) \Big] \nonumber \\
&+ \beta \Big[ p_s(1-\rho_c(x_s))\Delta v_s(b,x_s,c_{serve})\nonumber \\
&+ p_c \sum_{\tilde{x}_s=0}^{x_c}p_{\tilde{X}_s}(\tilde{x}_s)\Delta v_c(b,s,\tilde{x}_s,\alpha)\Big] \Big),
\label{eq:value_function_1}
\end{align}
where\\
$\mathcal{A}(b)$ is the two dimensional bid space $\mathcal{D} \times \mathcal{D}\cap[0,b+s/(1+\alpha)],$ \\
$\Delta v_s(b,x_s,c_{serve}) = v_{\rho}(b+x_s-c_{serve}) -v_{\rho}(b),$ and \\
$\Delta v_c(b,s,\tilde{x}_s,\alpha) = v_{\rho}(b+s-\tilde{x}_s-\alpha(\tilde{x}_s-b)^+)-v_{\rho}(b).$ Note that the latter two functions account for the change in value with a trade for a server and a client, respectively. Then, the space of possible value functions is
\begin{equation*}
\mathcal{V}=\braces{f:(\mathbb{R}_+\rightarrow \mathbb{R}):\|f\|_\infty<\infty}=L_\infty.
\end{equation*}
Define the Bellman operator $T_\rho$ on $L_\infty$ as below:
\begin{align}
&(T_\rho f)(b)= \beta f(b) + \max_{(x_s,x_c)\in\mathcal{A}(b)} \bigg( \Big(  p_s(1-\rho_c(x_s))(x_s \notag \\
&-c_{serve}) + p_c\big(\rho_s(x_c)(s-\mathbb{E}[\tilde{X}_s|\tilde{X}_s\leq x_c])- (1 \notag \displaybreak[0]\\
&-\rho_s(x_c))c_{lose} \big) \Big) + \beta \Big[ p_s(1-\rho_c(x_s))\Delta f_s(b,x_s,c_{serve}) \notag \\
&+ p_c \sum_{\tilde{x}_s=0}^{x_c}p_{\tilde{X}_s}(\tilde{x}_s)\Delta f_c(b,s,\tilde{x}_s,\alpha)\Big] \bigg)
\label{bell_op}
\end{align}
where
$\Delta f_s(b,x_s,c_{serve}) = f(b+x_s-c_{serve}) -f(b),$ and
$\Delta f_c(b,s,\tilde{x}_s,\alpha) = f(b+s-\tilde{x}_s-\alpha(\tilde{x}_s-b)^+)-f(b).$

\subsection{Properties of the Value Function}
In order to characterize the best response policy, we need to derive some useful properties of the value function $v_{\rho}$.  We start by proving the convergence of value iteration of the Bellman operator $T_\rho(\cdot)$.  This follows immediately from classical results by in \cite{hernandez1999further}, as long as we can prove the following three lemmas. Define the transition kernel $\mathcal{Q}(B|b,(x_s,x_c))$ for non-empty Borel subset $B\subset\mathbb{R}_+$  by equation (\ref{eq:state_trans}) together with the regeneration. Note that given $x_s,x_c$, the probability of trade happening can be directly calculated through $\rho$.

\begin{lemma}\label{lem:vi_1}
For every state $b\in \mathbb{R}_+$,\\
1) There exists an effective bid space $\hat{\mathcal{A}}(b),$ which is compact; \\
2) The reward-per-stage is lower semi-continuous in $(x_s,x_c)$; \\
3) The function $\mu(b,x_s,x_c) := \mathbb{E}_{\mathcal{Q}} [u(B)|b,(x_s,x_c)]$ is continuous in $(x_s,x_c)\in \hat{\mathcal{A}}(b)$ for every function $u\in \mathcal{V}$.
\end{lemma}
\begin{IEEEproof}
The proof of 1) follows from showing the existence of upper bounds on bids for both client and server yielding $\hat{\mathcal{A}}(b).$   The reward-per-stage in (\ref{bell_op}) is defined as  $c(b,(x_s,x_c))\triangleq p_s(1-\rho_c(x_s))(x_s-c_{serve}) + p_c \big(\rho_s(x_c)(s-\mathbb{E}[\tilde{X}_s|\tilde{X}_s\leq x_c])-(1-\rho_s(x_c))c_{lose} \big)$ and given $b,x_s,x_c$, the kernel $\mathcal{Q}$ is fully determined by $p_s,p_c,\rho,\Psi$.  The continuity of $\mathcal{Q}$ over the discrete topology of $\hat{\mathcal{A}}(b)$ is natural. Details of this proof are available in Appendix \ref{app_br}.
\end{IEEEproof}

\begin{lemma}\label{lem:vi_2}
There exist constants $\xi\geq 0$ and $\eta\geq 0$ with $1\leq \eta < 1/\beta$, and a function $w \geq 1$ s.t. for every state $b$ \\
1) $\sup_{\mathcal{A}(b)}|c(b,(x_s,x_c))|\leq \xi w(b)$; and \\
2) $\sup_{\mathcal{A}(b)}\mathbb{E}_{\mathcal{Q}} [w(B)|b,(x_s,x_c)]\leq \eta w(b).$
\end{lemma}
\begin{lemma}\label{lem:vi_3}
For every state $(b)$, the function $\omega (b,x_s,x_c):= \mathbb{E}_{\mathcal{Q}} [w(B)|b,(x_s,x_c)]$ is continuous in $(x_s,x_c)\in \hat{\mathcal{A}}(x)$.
\end{lemma}
\begin{IEEEproof}
Since $c_{serve}$, $s$ and $c_{lose}$ are fixed, $c(b,(x_s,x_c))$ is bounded. Then taking a bounded function $w$, with the continuity of $Q$, the results in Lemma \ref{lem:vi_2} and \ref{lem:vi_3} are straightforward.
\end{IEEEproof}
\begin{theorem} (Hernandez-Lerma \cite{hernandez1999further})\label{thm:vi_converge}
Given the belief $\rho_s,\rho_c$ and the corresponding p.m.f. $p_{\tilde{X}_s},p_{\tilde{X}_c}$ we have,\\
    1) There exists a $j\in\mathbb{N}$ such that $T_{\rho}^j:\mathcal{V}\rightarrow\mathcal{V}$ is a contraction mapping. Hence, there exists a unique $f_{\rho}^*\in\mathcal{V}$ such that $T_{\rho} f_{\rho}^*=f_{\rho}^*$, and for any $f\in\mathcal{V}$, $T^n_{\rho} f\rightarrow f_{\rho}^*$ as $n\rightarrow\infty$.\\
    2) The fixed point $f^*_{\rho}$ of operator $T_{\rho}$ is the unique solution to the Bellman equation, i.e., $f_{\rho}^*=v_{\rho}^*$.
\end{theorem}
\begin{lemma} \label{lem:inc_v}
$v^*_{\rho}(b)$ is monotonically increasing in $b$.
\end{lemma}
\begin{IEEEproof}
By Theorem \ref{thm:vi_converge}, we have proved $v_{\rho}$ converges to a unique fixed point $v_{\rho}^*$ over $T_{\rho}$. Thus, it is sufficient to prove that $T_{\rho}$ maintains the assumed monotonicity. Full details of the proof are presented in Appendix \ref{app_br}.
\end{IEEEproof}
\subsection{Best Response Policy Characterization}

As discussed in Section~\ref{sec:model}, our goal is to maximize server utilization from the system perspective, which is equivalent to maximize the expected trade ratio in the market.  Furthermore, the budget, which is defined through equation (\ref{eq:value_function}), increases through successful trade. These observations imply that we should characterize the best response policy not only from the single agent perspective, but also from the perspective of maximizing the expected trade ratio.  We will use this goal to motivate a specific family of equilibria for our problem. Given the four facts we discussed at the beginning of this section, we then show that for the best system performance a certain simpler class of bidding functions suffice.
\begin{lemma} \label{lem:single_price}
All servers bidding the same price within the clients' affordable range maximizes the expected trade ratio.
\end{lemma}
\begin{IEEEproof}
The proof follows from by comparing the trade ratios between the scenarios in which the server places multiple bids or a single bid.  Using the four facts, the corresponding client bid distributions can be further characterized.  Full details are available in Appendix \ref{app_br}.
\end{IEEEproof}

Motivated by Lemma~\ref{lem:single_price}, we characterize the best response policy by initializing the belief of server's bid distribution to be $p_{\tilde{X}_s}(k)=1$ for some fixed $k$, i.e. all servers bid the same price $k$.  For non-trivial behavior $k\geq c_{serve}$, but $k$ can be higher than $s$, though not by much, i.e. $s-k\geq c_{lose}$, otherwise, the trade will become worthless; see section \ref{sec:sim} for the latter.

\subsubsection{Client's Best Response}

Given the belief that all servers bid $k$, the value function of clients from (\ref{eq:value_function}) becomes
\begin{align}
& v_c(b)=\max_{x_c\in\mathcal{D}\cap[0,b+s/(1+\alpha)]} \Big( \bm{1}_{x_c \geq k}(\beta(v_{\rho}(b+s-k \nonumber\\
&\quad -\alpha(k-b)^+))+s-k)+\bm{1}_{x_c < k}(\beta v_{\rho}(b)-c_{lose}) \Big ) \nonumber
\end{align}
By Facts \ref{fact1} and \ref{fact3}, we conclude that the client will bid either $0$ or $k.$ If a client bids $k$, the trade will happen w.p. $1$, and will fail otherwise.  We define the following useful terms: \\
$v_{c\_win}(b) = \beta(v_{\rho}^*(b+s-k-\alpha(k-b)^+))+s-k$, \\
$v_{c\_lose}(b) = \beta v_{\rho}^*(b)-c_{lose}$,
$b_{c\_win} = b+s-k-\alpha(k-b)^+$, and
$b_{c\_lose}= b$.

Since the budget can never go negative, we have an upper limit on a client's bid of $b+s/(1+\alpha)$.  If $k$ lies out of this range, the client will simply bid $0$.  Now, from Lemma \ref{lem:inc_v}, $b_{c\_win} \geq b_{c\_lose}$ i.e. $b\geq ((1+\alpha)k-s)/\alpha=k-\frac{s-k}{\alpha}$ implies that if $v_{c\_win}(b) \geq v_{c\_lose}(b)$, then the client should bid $k$.  Thus, we have a lower bound on the bid as $0$, and the upper bound as $k$. The exact bidding strategy depends on the relationship between $v_{c\_win}(b)$ and $v_{c\_lose}(b)$.  A summary of the best responses of a client with budget $b$ is:
\begin{equation} \label{client_br}
x_c^*=  \\
\begin{aligned}
\begin{cases}
0 & b\in[0,k-\frac{s}{1+\alpha}) \\
0 \text{ if } v_{c\_win}(b) \leq v_{c\_lose}(b) & b\in[k-\frac{s}{1+\alpha},k-\frac{s-k}{\alpha}] \\
k \text{ if } v_{c\_win}(b) \geq v_{c\_lose}(b) & b\in[k-\frac{s}{1+\alpha},k-\frac{s-k}{\alpha}] \\
k & b\in(\frac{(1+\alpha)k-s}{\alpha}, \infty) \\
\end{cases}
\end{aligned}
\end{equation}

Note that when $k<s/(1+\alpha)$, all clients will bid $k,$ which is an extreme case of a ``cheap resource."  It implies the price one needs to pay is too low, as compared to the gain from the trade.  We further characterize the best response function $\theta_{c,\rho}(b)$ in the following Lemma.
\begin{lemma} \label{lem:finite_pieces}
$\theta_{c,\rho}(b)$ is piecewise constant on $[0,k-\frac{s-k}{\alpha}]$ with a finite number of constant intervals.
\end{lemma}
\begin{IEEEproof}
The proof follows by showing the difference $v_{c\_win}(b)-v_{c\_lose}(b)$ is of bounded total variation.  Full details are available in Appendix \ref{app_br}.
\end{IEEEproof}

\subsubsection{Server's Best Response}

Given the client's best response function, we observe that under certain circumstances, the client will bid either $0$ or $k$ based on her private state.  By Facts \ref{fact2} and \ref{fact4}, we conclude that the server will again bid either $0$ or $k$.  We can refine the server's belief about the client's bid distribution as $p_{\tilde{X}_c}=(z,1-z)$, where $z=\mathbb{P}(\tilde{X}_c=0)$.  Then
$ v_s(b) = (1-z)(\beta v_{\rho}(b+x_s-c_{serve})+x_s-c_{serve}) + z \beta v_{\rho}(b).$
By Lemma \ref{lem:inc_v}, for $\forall b\in\mathbb{R}_+$, we have $v_s(b)$ is monotonically increasing in $x_s$, when $x_s\leq k$.  Hence, all servers will bid $k$.

Given a feasible $k$ (which we refer to as a ``unified price'' for both clients and servers), the discussion above lends credence to the existence of an equilibrium over the simple set of beliefs given by $z$.  We will prove that such  Mean Field Equilibrium (MFE) indeed exists in section \ref{sec:mfe}.

\section{Mean Field Equilibrium}\label{sec:mfe}
The main result of this section is to show the existence of an MFE with under simple bidding strategies.  Given the unified price $k$ and the probability of bidding $0$ as a client, $z,$  the kernel of state transitions in (\ref{eq:value_function}) is well defined.  Denote the fixed point value function as $v_{z}^*$.  Taking the best response of client using (\ref{client_br}), we have the following budget transitions for a generic agent before she reveals her role:
\begin{equation} \label{markov_tran}
\begin{aligned}
b[t+1]=
\begin{cases}
    b[t]                     \text{ w.p. } \beta(p_s z + p_c \bm{1}_{b[t]\in B_0})                               \\
    b[t]+s-k-\alpha(k-b[t])^+\text{ w.p. } \beta p_c \bm{1}_{b[t]\in \mathbb{R}_+\setminus B_0} \\
    b[t]+k-c_{serve}         \text{ w.p. } \beta p_s (1-z)                                                       \\
    B_{init}                 \text{ w.p. } (1-\beta)\Psi(B_{init})
\end{cases}
\end{aligned}
\end{equation}
where, $B_0\subset \mathbb{R}_+,$ in which the agents bid $0$ as a client, and $\Psi$ is the probability measure of the agent regeneration process.  Set $B_{init} \subseteq \mathbb{R}_+$ is the set of possible budgets with regeneration.  When $b[t]$ lies on the boundaries of $B_0,$ by Lemma \ref{lem:finite_pieces}, a client is indifferent to bidding $0$ or $k$.  Also, the number of these boundary points in $B_0$ is finite, which leads to a Borel-null set in $B_0$.  Hence, w.l.o.g. adding these points by assuming the client will bid $0$ with some probability $p_{tie}$ and $k$ with the complementary probability will not alter the proofs in the rest of this section.

Observe that from (\ref{markov_tran}), given the current state $b[t]$, the next state $b[t+1]$ is independent of the rest of the history. Thus, the transition kernel above defines a Markov process for the budget, and we have following Lemma.
\begin{lemma} \label{lem:mc_pc}
The Markov chain $\{b[t]\}_{t=0}^{\infty}$ with transition kernel (\ref{markov_tran}) is positive recurrent and has a unique stationary distribution, $\pi_{z}$. Furthermore, given $z$, $\pi_{z}$ is absolutely continuous with respect to Lebesgue measure on $\mathbb{R}_+.$
\end{lemma}
\begin{IEEEproof}
The proof of the first statement follows by showing that the one-step transition function satisfies the Doeblin condition, then using the results in \cite[Chap. 12]{MeyTwe09}.  Then we derive the relationship between $\pi_{z}(B)$ and $\pi_{z}^{(\tau)}(B|b)$, where $\tau$ is the first regeneration time after $t=0$ and $b(0)=b$ to prove the second statement.  Details in Appendix \ref{app_mfe}.
\end{IEEEproof}

Combining the best response of each state $\theta_{z}$ and $\pi_{z},$ a new value $\tilde {z}$ can be calculated.  If $\tilde {z}=z$, then we say the system is at an MFE.  The main result of this section is to prove the following theorem, where $\theta_{c, z}(\cdot)$ is the set-valued (subset of $\{0,k\}$) function of client bids as a function of its budget.  More formally. $\theta_{c,z}:\mathbb{R}_+ \rightarrow \{0,k\}$ is the client's best response function which maps the budget to a binary choice of bids, either $0$ or $k$.  As mentioned earlier $\pi_{z}:\mathbb{R}_+ \rightarrow [0,1]$, defines the stationary distribution of the budget, which maps the state space to a probability measure. 
\begin{theorem} \label{thm:mfe}
Define $\gamma(z) \triangleq \pi_{z}(\theta_{c,z}^{-1}(0)),\forall z\in[0,1]$, where $\theta_{c,z}^{-1}(0)$ is the lower inverse of  $\theta_{c,z}(\cdot)$ at $0$.
There exists an MFE $(z,k,\theta_{z})$, such that $z = \gamma(z)$.
\end{theorem}

Define function $\gamma$ that maps the assumed belief $z$ to the resultant $\tilde {z} = \gamma(z)$ through the best-response dynamics.  To prove Theorem \ref{thm:mfe}, we need to show that $\gamma(\cdot)$ has a fixed point, i.e., $\gamma(z) = z$ for some $z \in [0,1]$.  Given the closed interval $[0,1]$ is compact and convex, by Brouwer Fixed Point Theorem, it is sufficient to show $\gamma$ is continuous.  However, the best response function $\theta_{c,z}(\cdot)$ is a set-valued function with a range containing all non-empty subsets of $\{0,k\}$ given the characterization in Section \ref{sec:br}.  We first show $\theta_{c,z}(\cdot)$ is upper hemicontinuous in $z$, then discuss the continuity of $\pi_{z}$.  Finally, we prove the single point inverse (lower inverse) of the upper hemicontinuous set-valued function $\theta_{c,z}(\cdot)$ is a subset that consists of finite number of continuous pieces, which leads to the continuity of $\gamma$ using the absolute continuity of $\pi_{z}$ with respect to Lebesgue measure from Lemma \ref{lem:mc_pc}.  The results with an outline of the proof are given below.


\begin{lemma} \label{lem:v_lip}
$v^*_{z}$ is Lipschitz continuous in $z$.
\end{lemma}
\begin{IEEEproof}
The proof follows using the properties of the contraction mapping $T^j_{z}$ in Theorem \ref{thm:vi_converge}. Full details are presented in Appendix \ref{app_mfe}.
\end{IEEEproof}

\begin{theorem}\label{thm:theta_con}
$\theta_{c,z}(\cdot)$ is upper hemicontinuous in $z$.
\end{theorem}
\begin{IEEEproof}
Given $z$ and $k$, we can rewrite $v^*_z$ in a different way such that it can be represented as a increasing piecewise linear convex function. The proof holds by applying Berge's Maximum Theorem. For full details see Appendix \ref{app_mfe}.
\end{IEEEproof}

\begin{theorem} \label{thm:pi_con}
$\pi_{z}$ is continuous in $z$.
\end{theorem}
\begin{IEEEproof}
The key idea of the proof is using Portmanteau Theorem to show for any uniform converging sequence $z_n\rightarrow z$ and any open set $B$, $\liminf_{n\rightarrow\infty}\pi_{z_n}(B)\geq\pi_{z}(B)$. Details of this proof are presented in Appendix \ref{app_mfe}.
\end{IEEEproof}

\begin{theorem}\label{thm:gamma_con}
$\gamma(z) \triangleq \pi_{z}(\theta_{c,z}^{-1}(0))$ is continuous in $z$.
\end{theorem}
\begin{IEEEproof}
We prove this by showing that $\theta_{c,z}^{-1}(0)$ is a continuity set for $\pi_{z}$. See Appendix \ref{app_mfe} for details.
\end{IEEEproof}


\section{Simulation}\label{sec:sim}
\subsection{Statistics of a Computing Cluster}
As mentioned in Section \ref{sec:intro}, one of our candidate applications is sharing computational resources. We consider the statistics of three computing clusters at the authors' institution over two weeks: the hourly mean utilizations $68.5\%$, $54.8\%$, $45.4\%$ and median utilizations $70\%$, $55\%$ and $47.5\%$. We observe that the mean is similar to the median in all three cases, which implies half of the time a cluster has  a high load, and  in the other half, it has available capacity.    While it is difficult to obtain precise values of the other parameters for our model for this application, we can use it to compare equilibria for different parameter settings.  We will conduct a more realistic case study on energy trading in the next section.  



\subsection{Monte Carlo Simulations}
We conduct Monte Carlo simulations of three versions of our system over one million virtual agents (cluster owners). The three models are designed as following:\\
\textbf{Hard Model:} Overdraft as a client is not allowed. The budget $b$ is a hard constraint of the client's bidding space.\\
\textbf{Bank Model:} Here, the client can overdraw the budget to a certain extent such that a non-negative post-trade budget is guaranteed. The amount overdrawn comes from an external source, i.e., a bank, and the client has to pay both the principal and interest back.\\
\textbf{Loan Model:} The server takes on the role of the bank. It implies the server and the client agree to mining the surplus $s$ together through the trade. The server obtains additional revenue beyond the payment $k$ in doing so. In this case, the server's expected payoff is higher than the bank model.

We believe that similar proofs of the existence and nature of equilibrium apply to all three cases above and not just the bank model.  Before presenting simulation results, we first introduce the parameter settings. We use a regeneration factor $\beta=0.98$; we assume that agents have an equal probability being clients and servers, i.e., $p_s=p_c=0.5$; $\alpha=1.1,$ which means that the interest rate is set to be $10\%$; $s=8$ and $c_{serve}=6$ make the trade generate reasonable amount of value; $c_{lose}=0.5$ captures the client disappointment when the trade fails; $\Psi(B_{init})=U[0,5]$, new agents come with limited amount of budget, which helps us better analyze the differences among three models; and price $k=7$ in order to balance the benefits between servers and clients through the trade. Later in this section, we will show how different prices and initial budgets affect the equilibrium trade ratio in the bank model.

\begin{figure*}[ht]
\begin{minipage}{.3\linewidth}
\centering
\hspace{-0.15in}
\includegraphics[width=1\linewidth]{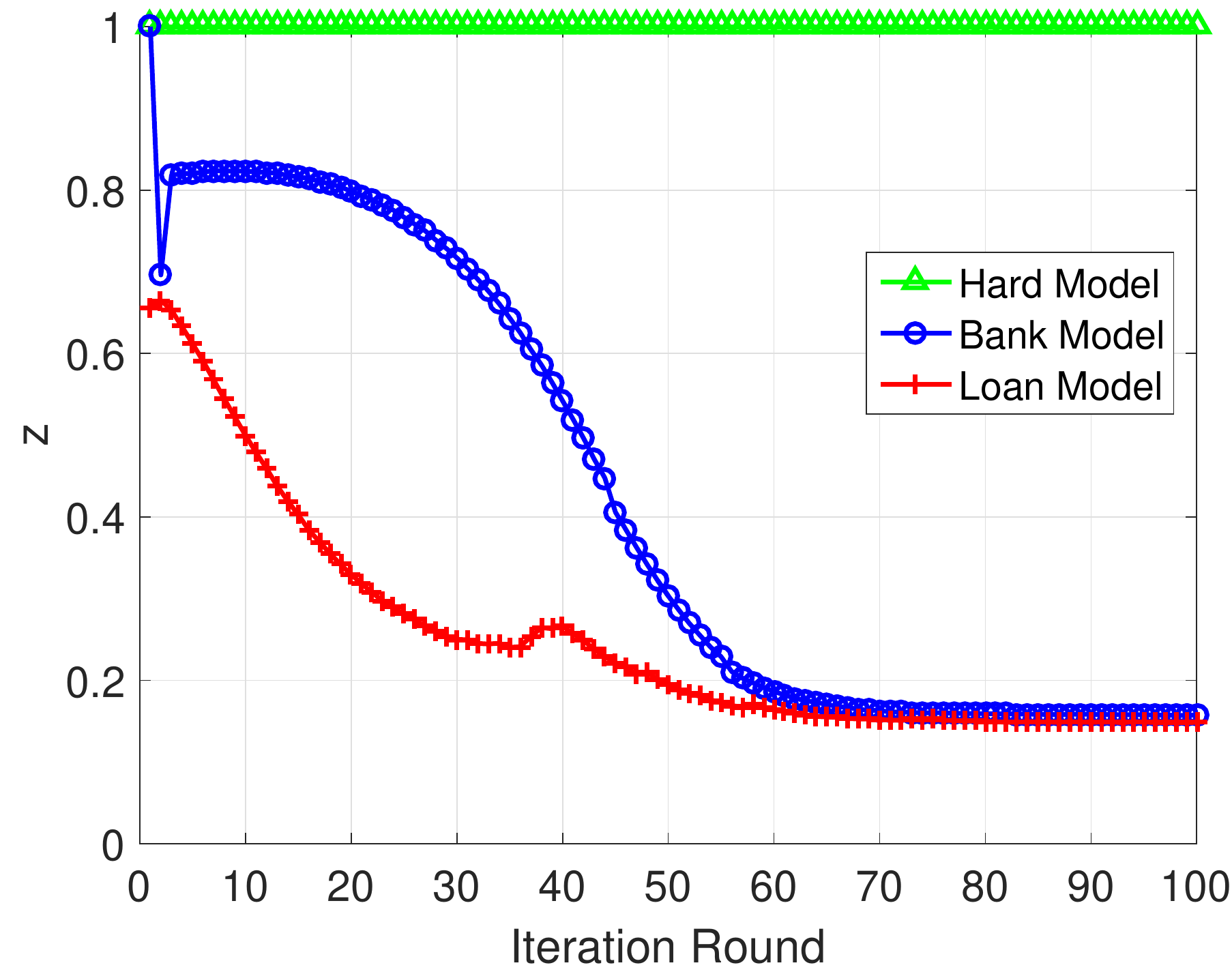}
\caption{Convergence of the belief, $z$}
\label{fig.bid_converge}
\end{minipage}\hfill
\begin{minipage}{.3\linewidth}
\centering
\hspace{-0.15in}
\includegraphics[width=1\linewidth]{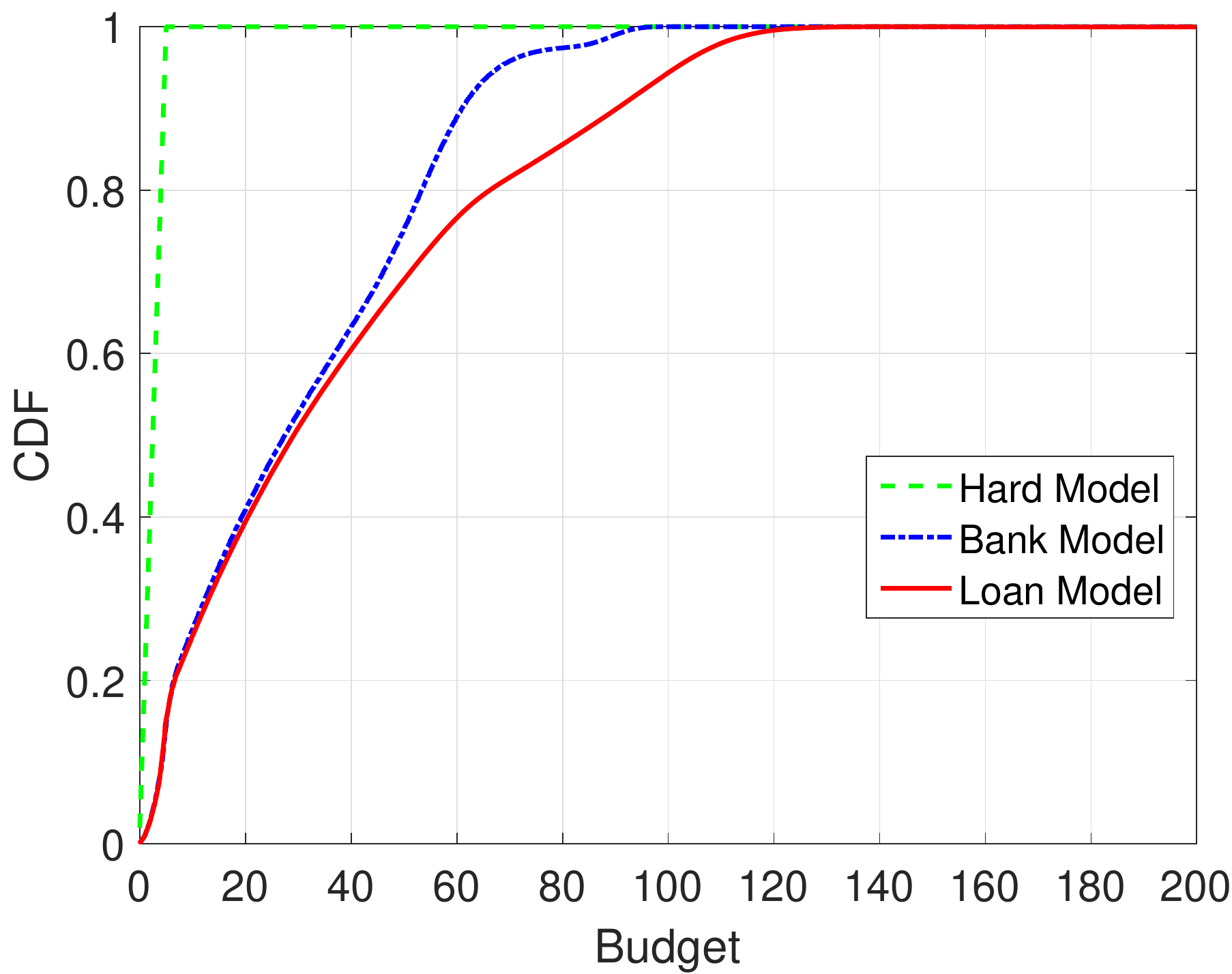}
\caption{CDF of budget at MFE}
\label{fig.b_dist}
\end{minipage}\hfill
\begin{minipage}{.3\linewidth}
\centering
\hspace{-0.15in}
\includegraphics[width=1\linewidth]{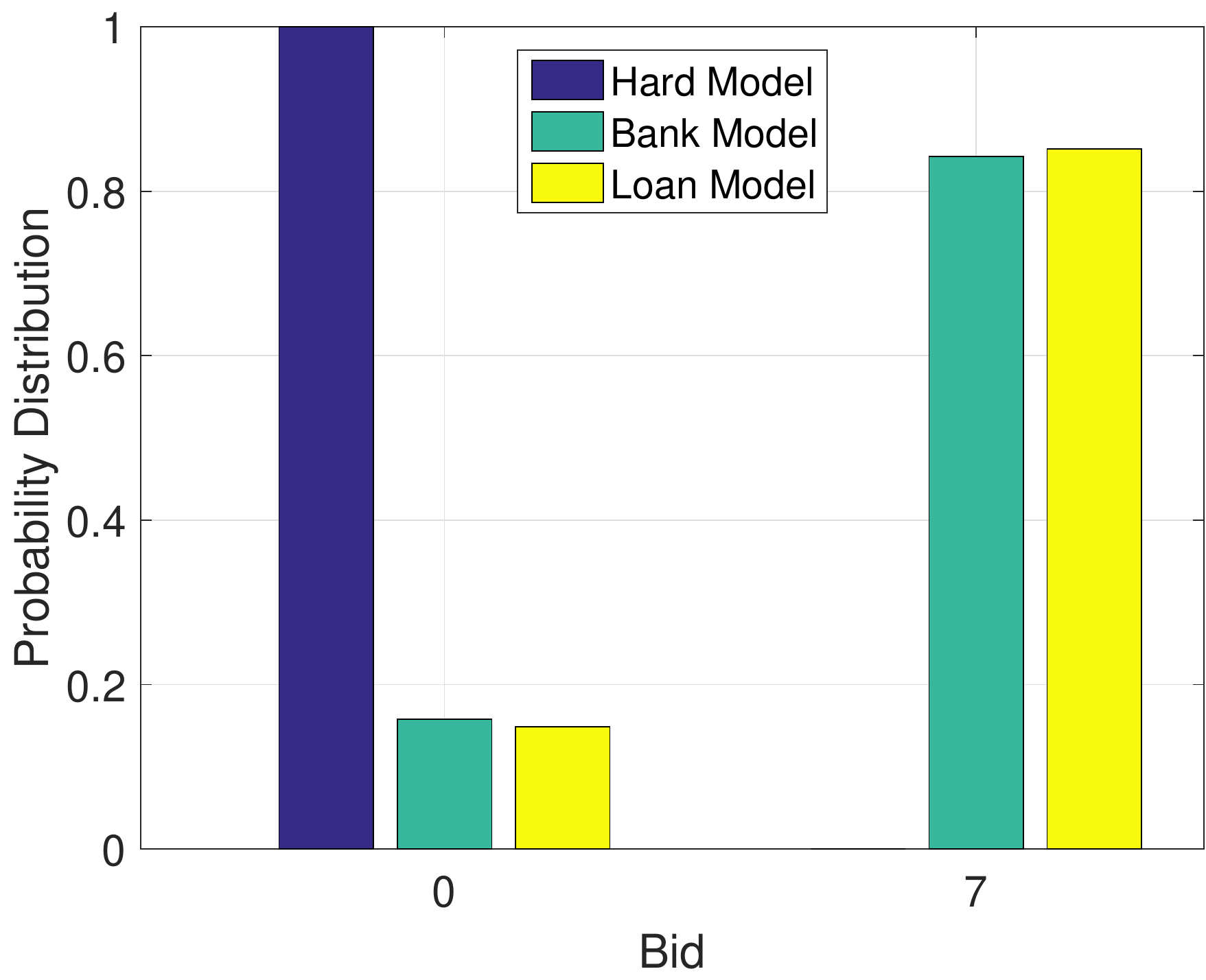}
\caption{Client bid distribution at MFE}
\label{fig.bid_bar}
\end{minipage}\hfill
\begin{minipage}{.3\linewidth}
\centering
\includegraphics[width=1\linewidth]{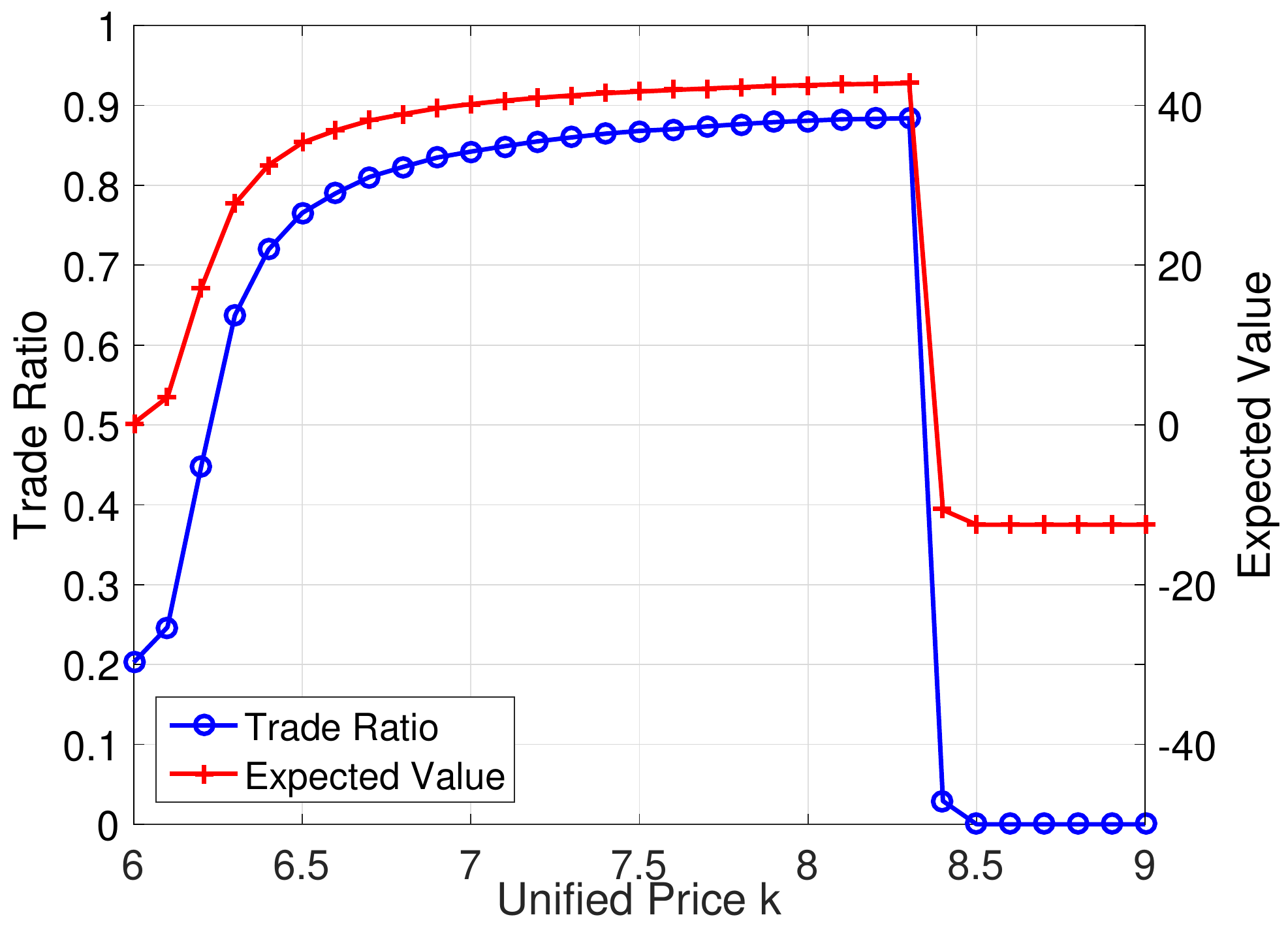}
\caption{Trade ratio and expected value vs. $k$, $\Psi(B_{init})=U[0,5]$}
\label{fig.k_tr_1}
\end{minipage}\hfill
\begin{minipage}{.3\linewidth}
\centering
\includegraphics[width=1\linewidth]{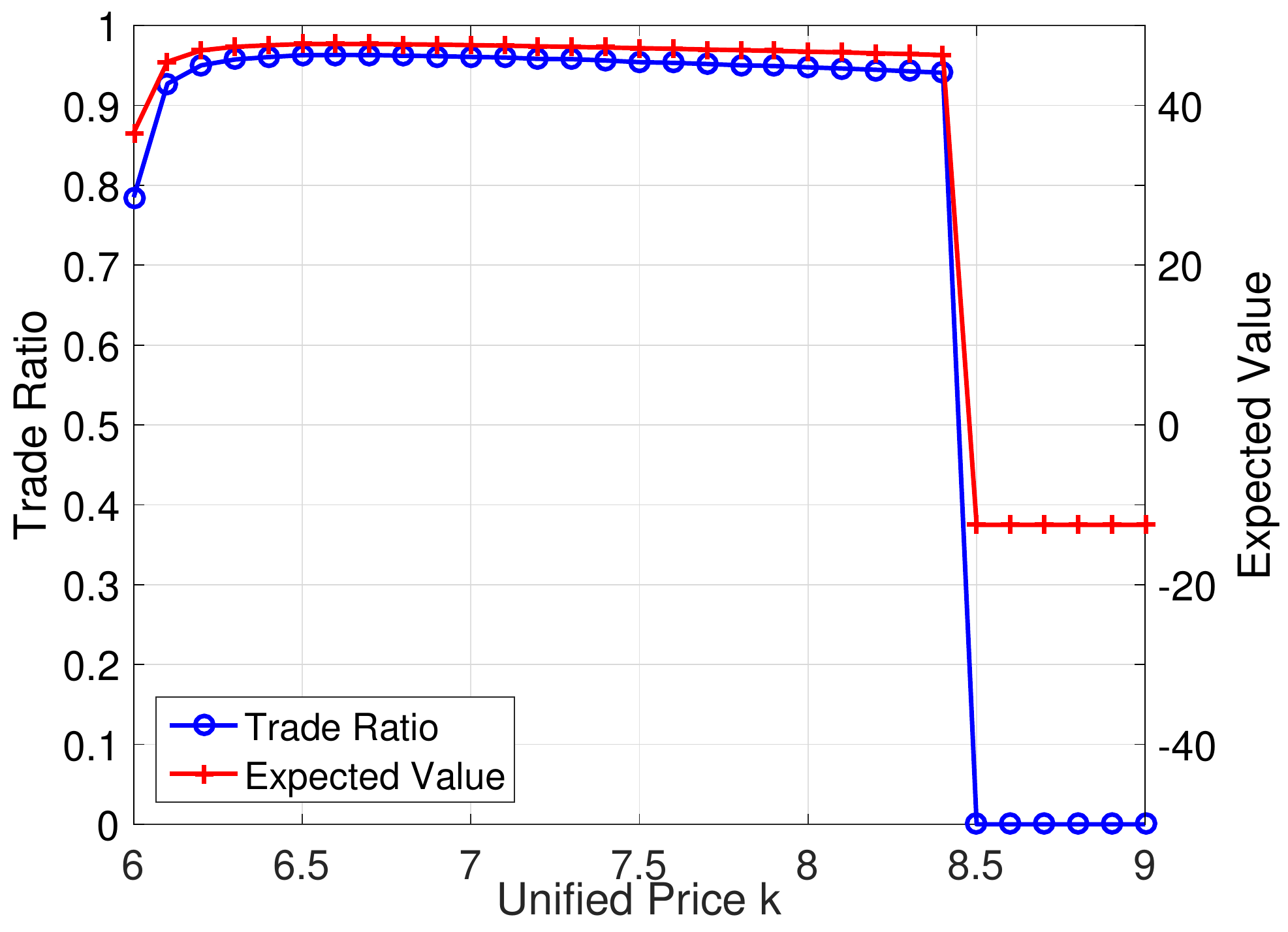}
\caption{Trade ratio and expected value vs. $k$, $\Psi(B_{init})=U[3,8]$}
\label{fig.k_tr_2}
\end{minipage}\hfill
\begin{minipage}{.3\linewidth}
\centering
\includegraphics[width=1\linewidth]{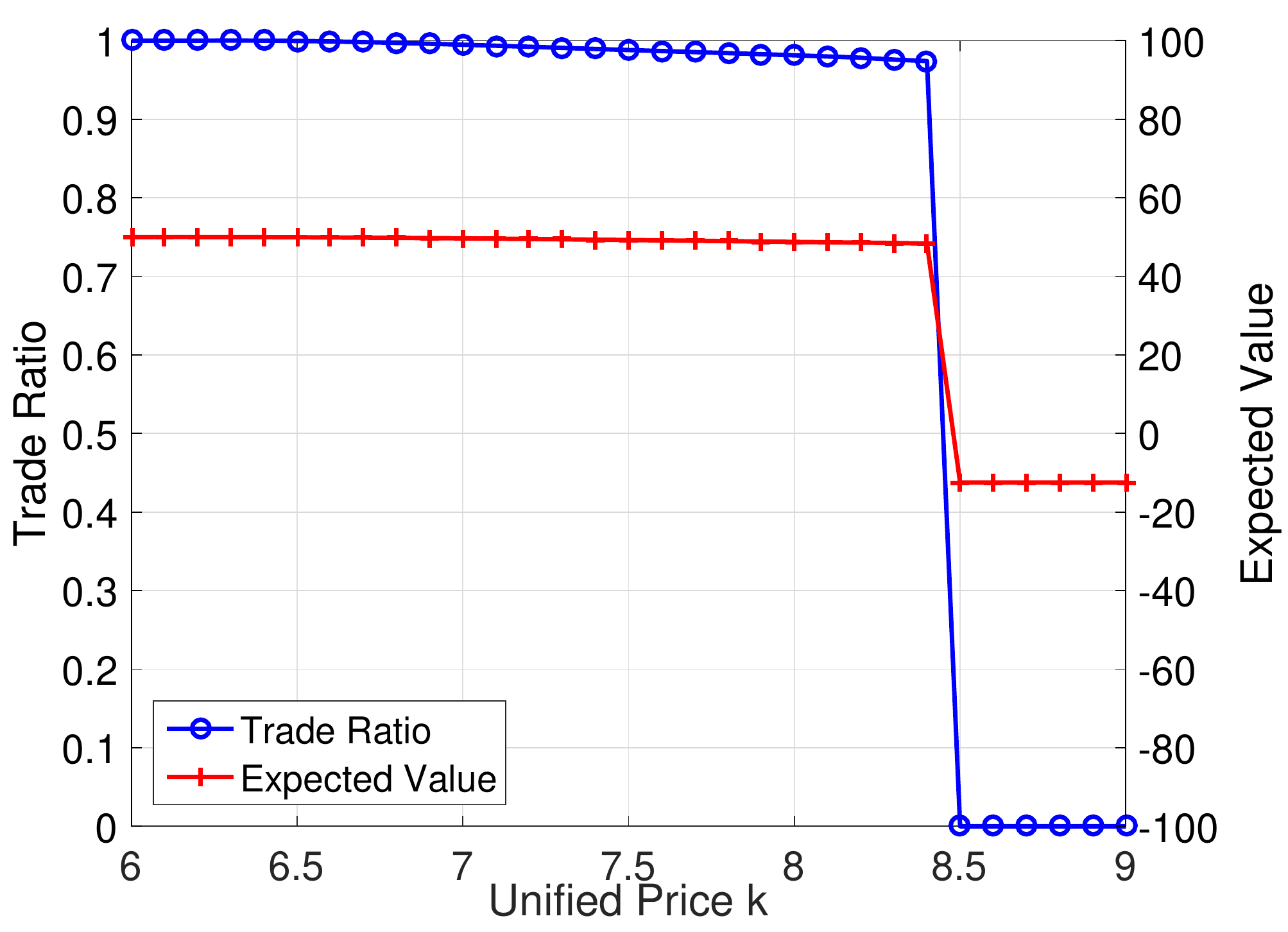}
\caption{Trade ratio and expected value vs. $k$, $\Psi(B_{init})=U[5,10]$}
\label{fig.k_tr_3}
\end{minipage}\hfill
\begin{minipage}{.3\linewidth}
\centering
\includegraphics[width=1\linewidth]{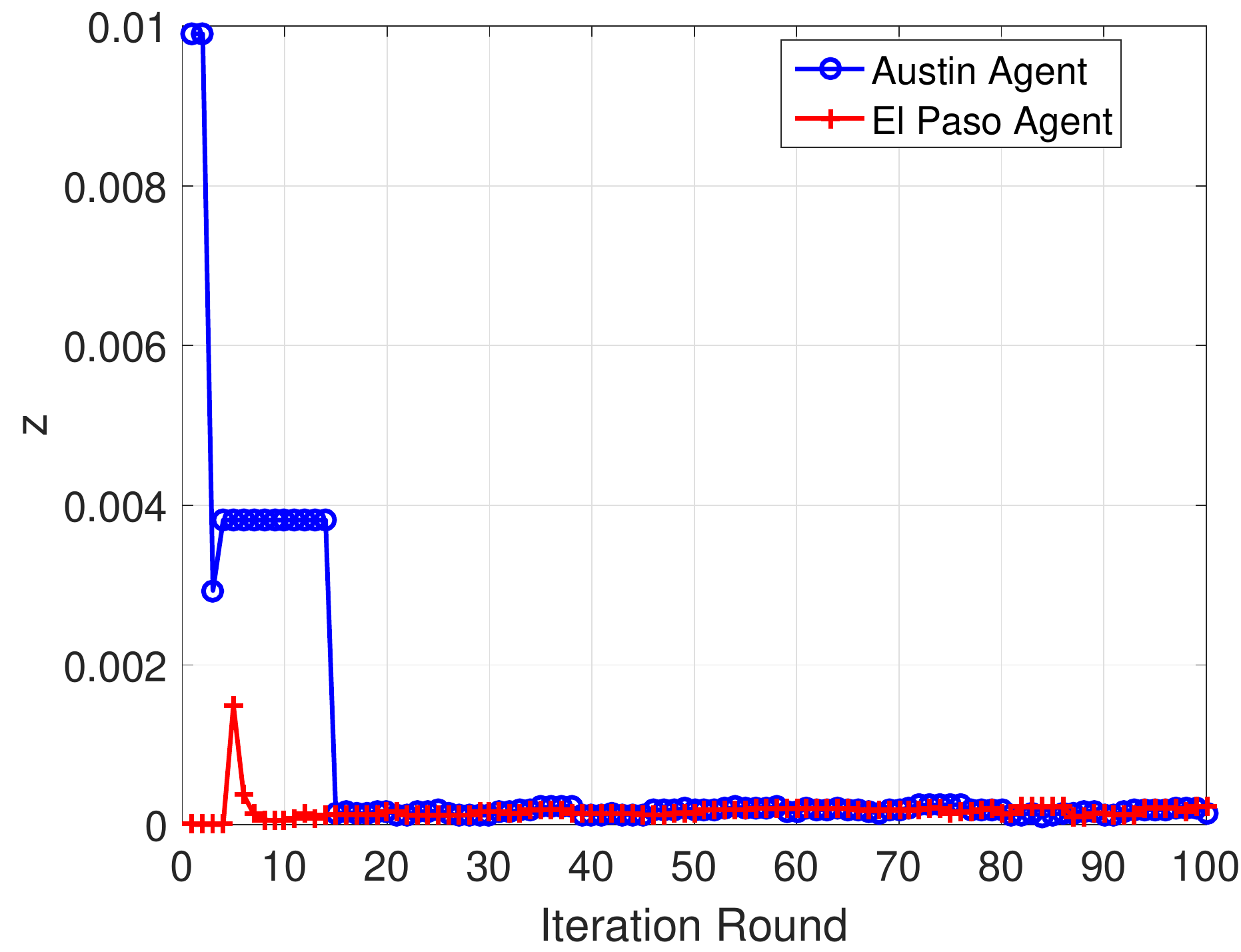}
\caption{Convergence of the Coupling Belief}
\label{fig.case_con}
\end{minipage}\hfill
\begin{minipage}{.3\linewidth}
\centering
\includegraphics[width=1\linewidth]{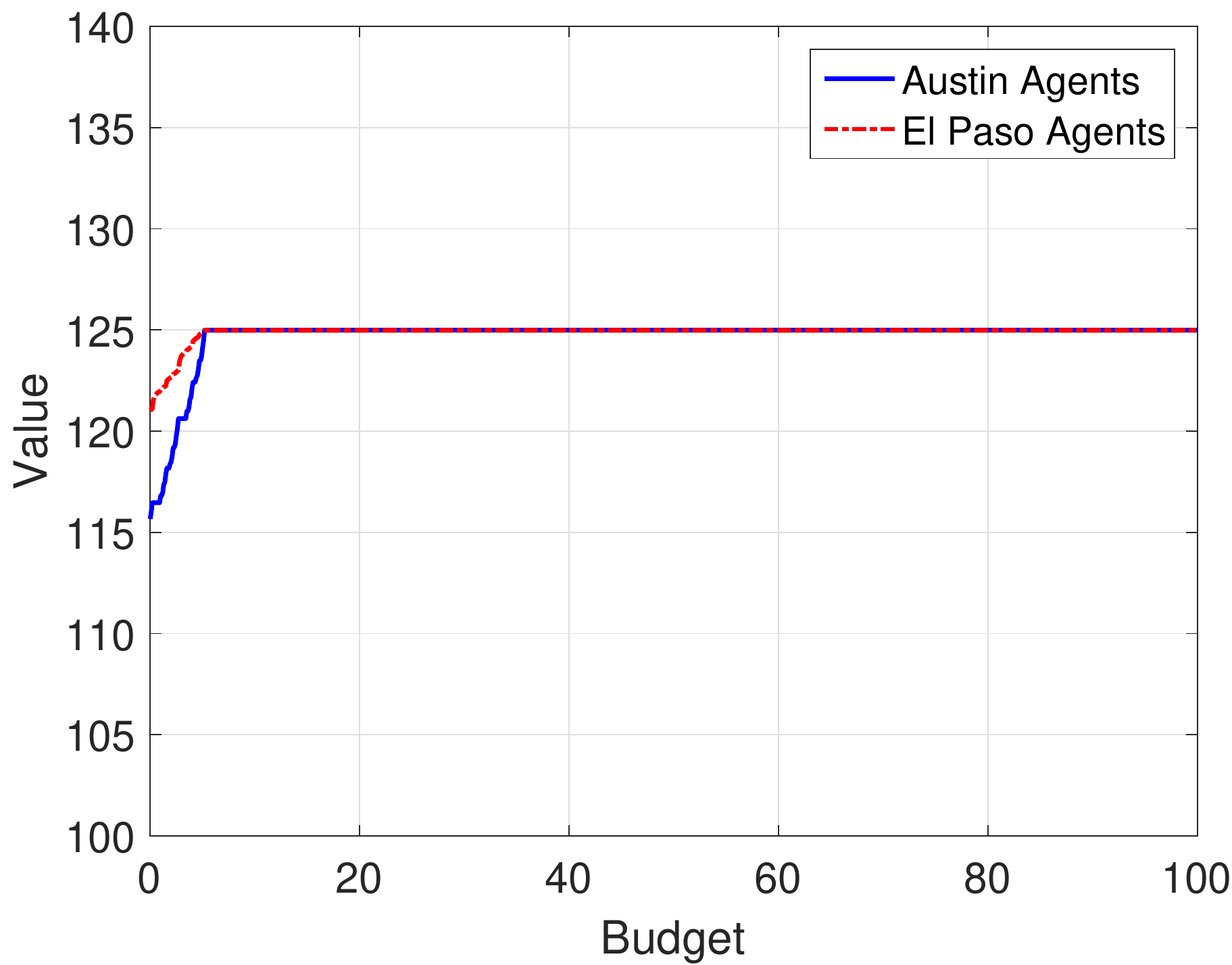}
\caption{MFE Value Function}
\label{fig.case_v}
\end{minipage}\hfill
\begin{minipage}{.3\linewidth}
\centering
\includegraphics[width=1\linewidth]{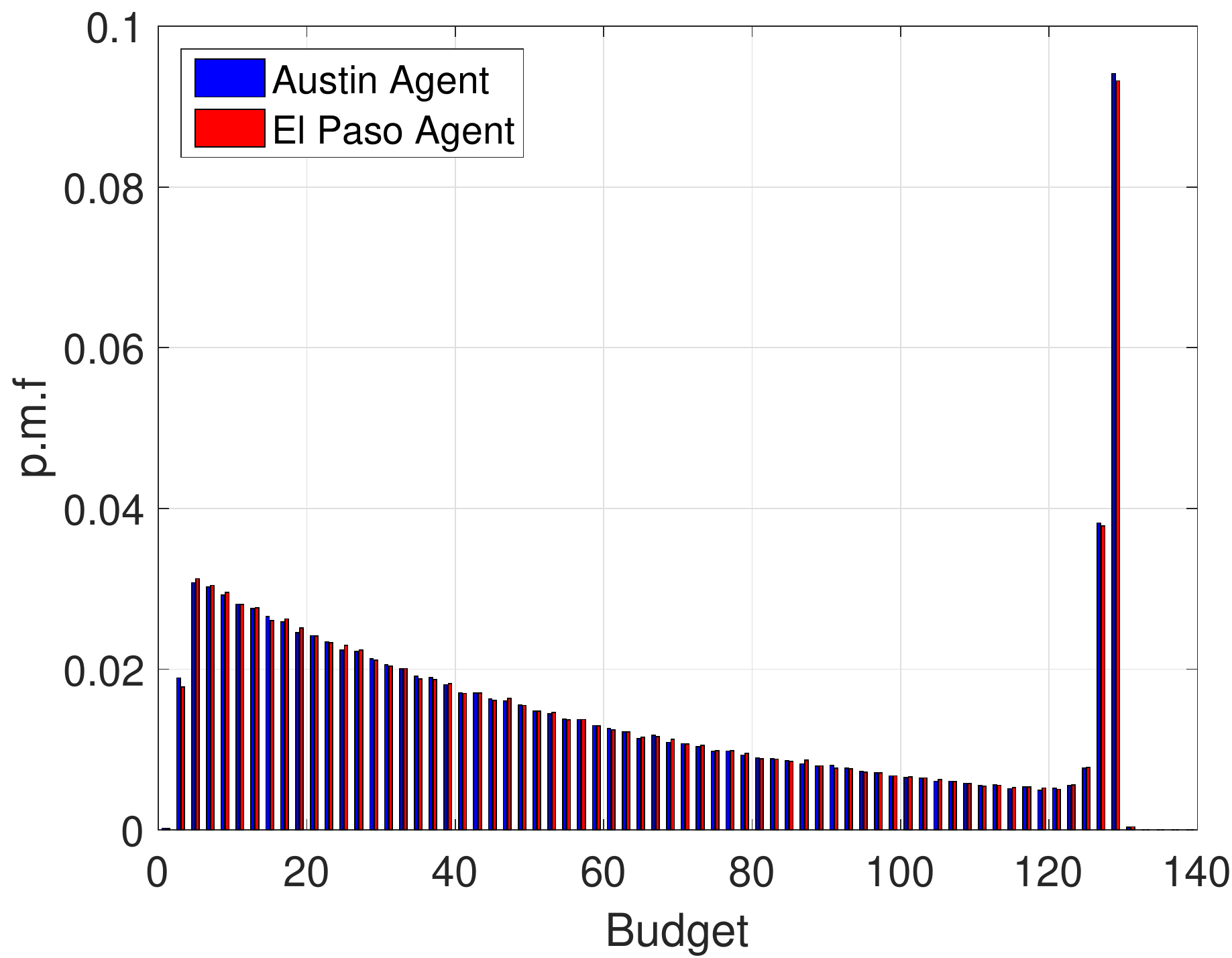}
\caption{MFE Budget Distribution}
\label{fig.case_bd}
\end{minipage}\hfill
\end{figure*}
Fig.~\ref{fig.bid_converge} shows the convergence of the MFE belief, $z$. In the hard constraint model, clients cannot afford the price $k=7,$ so all clients bid $0$ and the system freezes. However, in the bank model and loan model, the system gradually ramps up through different borrowing mechanisms and $z$ dramatically reduces when the system attains more and more wealth through successful trades.  Fig.~\ref{fig.b_dist} presents the CDFs of budget at MFE across the three models, which indicates agents are wealthier in the loan model than in the bank model at MFE (as the bank extracts some of the surplus), while the budget distribution is similar to the initial values in the hard constraint model. 
Fig.~\ref{fig.bid_bar} shows the binary bid distributions of clients at MFE, which verifies the results in Section \ref{sec:br}.
\begin{table}[h]
\caption{Trade Ratio and Expected Value}
\label{tb:tr_ev}
\centering
\begin{tabular}{|c|c|c|c|c|c|c|}
\hline
Initial Budget&\multicolumn{3}{|c|}{$\Psi(B_{init})=U[0,5]$} & \multicolumn{3}{|c|}{$\Psi(B_{init})=U[5,10]$} \\
\hline
Model & Hard & Bank & Loan & Hard & Bank & Loan\\
\hline
Trade Ratio & 0\%& 84.3\%& 85.2\% & 97.7\%& 99.4\%& 99.5\%\\
\hline
Value & -12.49& 40.14& 41.74& 48.53& 49.6& 49.7\\
\hline
\end{tabular}
\end{table}
We further evaluate two important statistics, namely the trade ratio and the expected value in Table \ref{tb:tr_ev}, where the expected value is calculated according to  $\mathbb{E}_{\Psi(B_{init})}[v^*_{z(MFE)}]$. 
As we mentioned earlier, the higher trade ratio the market system achieves, the higher resource utilization it ends up with.  Given the unified bid of server and binary bids of client, the expected trade ratio is captured by $1-z$ at MFE.  The empirical trade ratio among the one million agents matches this quantity in all cases. Intuitively, a higher trade ratio implies a higher expected value, which is verified in Table \ref{tb:tr_ev}.  For the sake of comparison, we also append the results of sufficient initial budget case, i.e. $\Psi(B_{init})=U[5,10]$, in Table \ref{tb:tr_ev} as well.  We observe that with a sufficient initial budget, the hard constraint model also achieves a reasonably high trade ratio at MFE, and the statistics of bank model catches up with the loan model and reaches 100\% trade ratio.

In the simulations above, we set the unified price $k=7$ and the initial budget, $B_{init}$, to be uniform distributed in $[0,5]$. Next, we will show how these two attributes affect the trade ratio of the market system.  Fig.~\ref{fig.k_tr_1}, \ref{fig.k_tr_2} and \ref{fig.k_tr_3} illustrate the trade ratio and the corresponding expected value versus different unified prices with different initial budget distributions under the bank model. Recall that $s=8$ and $c_{serve}=6$, which means that the reasonable range of $k$ lies in $[6,8+\epsilon]$, given the client is allowed to overdraw his budget to some extent.  We observe that for $B_{init}$ in $[0,5]$, the trade ratio increases in $k$; for $B_{init}$ in $[3,8]$, the trade ratio first increases then decreases; for $B_{init}$ in $[5,10]$, the trade ratio decreases in $k$.

An intuitive explanation of these results is as follows.  When most agents have a sufficient budget, lower price yields higher trade ratio. However, when most agents have a limited budget, it is better to set higher prices to aggregate the wealth at fewer agents, which then can obtain service without needing bank assistance (which reduces the total wealth of the system).  

\section{Case Study: Photovoltaic Market}\label{sec:case}
In this section, we consider a PV energy sharing market discussed in Section \ref{sec:intro}, in which the agents could be householders or small business owners.  During the hours with high sunshine, the system can supply the electricity consumption of a single agent.  Moreover, it even generates extra energy that can be  fed back to the grid.  However, in rainy or overcast weather, solar panels can only produce 10\%-25\% of their rated capacity \cite{cloudy_solar}.  This creates the opportunity to share the extra solar energy between locations with good and bad weather.

In this case study, we pick two big cities in Texas---Austin and El Paso---as an example. Note that  we choose the two locations far apart from each other to reduce the weather correlations.  Also, it is easier to share electricity energy over the existing grid facilities for locations in the same state.  Collecting the hourly historical weather data of the two cities in $2016$ from \cite{weather_under}, we found that approximately 44\% of the time, both cities have good weather and about 11\% of the time both have bad weather (hence do not need the market).   

We are interested in situations in which the two cities have different weather so that they could share extra solar energy between each other.  It turns out that the probabilities these events are $0.28$(Austin-bad, El Paso-good)  and $0.17$ (Austin-good, El Paso-bad).  Normalizing these values, we obtain two type of agents: Austin agents that are clients about 60\% of the time and are servers 40\% of the time, and El Paso agents who are exactly the opposite.  Given the heterogeneous types of agents, we have a slightly different set up than in the cloud computing case, in which all agents were homogeneous.  Also, given the regional weather effects, all Austin agents are of one type, while all El Paso agents are of one type.
We will show the convergence of the coupling beliefs and the consistency of budget distributions across agent types under the chosen market price, $k$.  The setting resembles the one in \cite{MHuang2010MFP}, with multiple agent types that undergo exogenous changes.

We choose parameters for our anlysis based on readily available data.  As we discussed above, $p_s$ and $p_c$ for Austin (El Paso) agents are 40\%(60\%) and 60\%(40\%) respectively. The average solar system size in the U.S. is $5kw$ \cite{avg_solar}.  Given that the average electricity usage is roughly $2.5kwh$ per hour in daytime \cite{elec_usage}, a server is able to provide $2.5kwh$ extra energy with one hour full sunshine, which is sufficient to supply a typical client. Given the average electricity price to be $10$ cents/kwh, we set the value of surplus $s=10$. However, an accurate unit cost for rooftop solar energy is not well established, since it varies by the installation fee, the maintenance fee, the government subsidy, etc., and we use $5$ cents/kwh as a conservative estimate \cite{solar_cost}, which yields $c_{serve}=5.$   For simplicity of simulation, we take the currency unit in our simulation as $2.5$ cents (so that one unit of service is traded), and multiply back by this amount after completion.  We choose the unified price $k=7.5$ to balance the benefits of the trade between clients and servers. A sufficient budget initialization $\Psi(B_{init})=U[5,10]$ is used.

Fig.~\ref{fig.case_con} shows the convergence of the coupled beliefs. We see that El Paso clients get involved in trades faster than Austin clients, since the higher probability of being servers makes them get wealthy faster. Also, the higher probability of being a client for Austin agents makes their value of budget lower, which is shown in Fig.~\ref{fig.case_v}.   Fig.~\ref{fig.case_bd} illustrates the consistency of budget distributions between the two agent types.  With the balanced price chosen, the consistency of budget distributions implies a high trade ratio.  Indeed, the trade ratio turns out to be $99.9\%$, which matches our discussion of the sufficient budget case in Section \ref{sec:sim} as well.

Finally, we evaluate our market system using weather traces in the year $2016$.  We conservatively define daytime to be the interval between one hour after sunrise to one hour before sunset, and found that in $1115$ time periods (hours) an Austin agent is a client whereas in $692$ time periods, an El Paso agent is a client.  We now calculate the savings for the agents participating in our proposed market and compare it with the case that one uses Net Metering.  One potentially saves $(1115*(10-7.5)+692*(7.5-5))*2.5/100*99.9\%\approx\$112.8/year$ through our market.  Note that the price $k$ balances the benefits of the trade, so the savings are consistent between different types of agents.  However, as a Net Meter user, the grid usually pays at the rate $5$\textcent/kwh \cite{solar_rate}, which gives zero profit as a server and encounters deficits as a client, i.e. $-1115*10/100=-\$111.5/year$ for an Austin agent and $-692*10/100=-\$69.2/year$ for an El Paso agent.

\section{Conclusion}\label{sec:conc}
In this paper, we considered the problem of market equilibria that arise in sharing economies where agents change roles frequently from provider (server)  to consumer (client).  We developed a model of bilateral trade under which consumers and providers are matched randomly with each other.  Under the MFG setting, we showed that the MFE consists of a single price bid by both client and server.   We conducted numerical evaluations to study the effects of different equilibrium prices on trade ratios, and showed in a case study that significant savings are possible in a rooftop PV setting.

\bibliographystyle{IEEEtran}
\bibliography{IEEEfull,small_scale_markets}  

\appendix
\subsection{Proofs in Section \ref{sec:br}} \label{app_br}

\textbf{Proof of Lemma \ref{lem:vi_1}}
Recall the definition of $\mathcal{A}(b):\mathcal{D} \times \mathcal{D}\cap[0,b+s/(1+\alpha)]$. We claim the effective bid of a server, $x_s$, lies in a compact set with an upper bound $\bar{x}_s$, which follows from the argument that no server will bid above $\bar{x}_s<\infty$ such that the expected return is below $c_{serve}$ (since this has to be paid in every time-frame). By (\ref{eq:value_function}), when a server bids $x_s$ we have only the clients with budget $b\geq x_s-s/(1+\alpha)$ will respond so that the expected return is given by $x_s \rho_c(x_s-s/(1+\alpha))$. However, all budget distributions in our system are stochastically dominated by the distribution obtained by transferring all wealth $s-c_{serve}$ to both the client or server in every time period. Note that the initial budget is given by the regeneration distribution that has support $B_{init}$. We will assume that $B_{init}$ is bounded with upper-bound $\bar{b}_{init}$. Given the lifetime of an agent in the system is geometrically distributed with parameter $1-\beta$, we have
\begin{align*}
\rho_c\left(x_s-\frac{s}{1+\alpha}\right) \leq \beta^{x_s-s/(1+\alpha)-\bar{b}_{init}}.
\end{align*}
Thus, we have the expected return of the server
\begin{align*}
x_s\rho_c(x_s-s/(1+\alpha))& \leq x_s \beta^{x_s-s/(1+\alpha)-\bar{b}_{init}}, \text{ and}\\
\lim_{x_s\rightarrow\infty} x_s\rho_c(x_s-s/(1+\alpha)) & \leq  \lim_{x_s\rightarrow\infty} x_s \beta^{x_s-s/(1+\alpha)-\bar{b}_{init}} \\
& = 0.
\end{align*}
Define the following
\begin{align*}
R&\triangleq \max_{x_s\geq 0}  x_s \beta^{x_s-s/(1+\alpha)\bar{b}_{init}}.
\end{align*}
We will assume that the parameters $(\beta, s, \alpha)$ are chosen such that $R\geq c_{serve}$. Under this assumption we set $\bar{x}_s$ to be the largest root of the transcendental equation $$c_{serve}=x \beta^{x-s/(1+\alpha)-\bar{b}_{init}}.$$
Meanwhile, as a client, given a finite budget $b$, the closed interval $[0,b+s/(1+\alpha)]$ is compact. Hence the effective action space lies in a compact set $\hat{\mathcal{A}}(b)\triangleq \mathcal{D}\cap[0,\bar{x}_s] \times \mathcal{D}\cap[0,\bar{x}_c] $, where $\bar{x}_c \triangleq b+s/(1+\alpha)$.

We define the reward per stage as
\begin{equation*}
\begin{aligned}
&c(b,(x_s,x_c))\\
& \triangleq p_s(1-\rho_c(x_s))(x_s-c_{serve}) \\
&\quad +p_c \big(\rho_s(x_c)(s-\mathbb{E}[\tilde{X}_s|\tilde{X}_s\leq x_c])-(1-\rho_s(x_c))c_{lose} \big).
\end{aligned}
\end{equation*}
Since $c_{serve}$, $s$ and $c_{lose}$ are constants, we have $c(b,(x_s,x_c))$ is bounded and continuous in $(x_s,x_c)$. Finally, the third result follows from the continuity of the transition kernel $\mathcal{Q}$ over discrete topology of $\hat{\mathcal{A}}(b)$.

\textbf{Proof of Lemma \ref{lem:inc_v}}
Suppose $f_n$ is monotonically increasing and $x_s^*,x_c^*$ maximize $T_{\rho}f_n(b) = f_{n+1}(b)$. Let $b'>b$ we have
\begin{align*}
&f_{n+1}(b') \\
&\geq p_s \bigg( \mathbb{E}_{\rho}[\bm{1}_{\tilde{x}_c \geq x_s^*}(\beta f_n(b'+x_s-c_{serve})+x_s-c_{serve})\\
&+\bm{1}_{\tilde{x}_c < x_s^*}\beta f(b')] \bigg)+ p_c \bigg ( \mathbb{E}_{\rho}[\bm{1}_{x_c^* \geq \tilde{x}_s}(\beta f_n(b'+s-\tilde{x}_s\\
&-\alpha(\tilde{x}_s-b')^+)+s-\tilde{x}_s)+\bm{1}_{x_c^* < \tilde{x}_s}(\beta f_n(b')-c_{lose})] \bigg )\\
&\geq p_s \bigg( \mathbb{E}_{\rho}[\bm{1}_{\tilde{x}_c \geq x_s^*}(\beta f_n(b+x_s-c_{serve})+x_s-c_{serve})\\
&+\bm{1}_{\tilde{x}_c < x_s^*}\beta f(b)] \bigg)+ p_c \bigg ( \mathbb{E}_{\rho}[\bm{1}_{x_c^* \geq \tilde{x}_s}(\beta f_n(b+s-\tilde{x}_s\\
&-\alpha(\tilde{x}_s-b)^+)+s-\tilde{x}_s)+\bm{1}_{x_c^* < \tilde{x}_s}(\beta f_n(b)-c_{lose})] \bigg )\\
&= f_{n+1}(b)
\end{align*}

\textbf{Proof of Lemma \ref{lem:single_price}}
Denote the expected trade ratio variable as $\kappa$. Consider the case in which servers bid multiple values, w.l.o.g., we assume $p_{\tilde{X}_s}=(p_{\tilde{X}_s}(k_1),p_{\tilde{X}_s}(k_2))$, where $c_{serve}<k_1<k_2<s$ and all clients can afford $k_2$. By the four facts, we have the client will bid either $k_1$ or $k_2$. We denote the probabilities of clients placing such bids as $p_{\tilde{X}_c}=(p_{\tilde{X}_c}(k_1),p_{\tilde{X}_c}(k_2))$. This gives us
\begin{equation*}
\begin{aligned}
\kappa =& p_{\tilde{X}_c}(k_1)p_{\tilde{X}_s}(k_1)+p_{\tilde{X}_c}(k_2)(p_{\tilde{X}_s}(k_1)+p_{\tilde{X}_s}(k_2))\\
=&p_{\tilde{X}_c}(k_1)p_{\tilde{X}_s}(k_1)+p_{\tilde{X}_c}(k_2)
\end{aligned}
\end{equation*}
Now if the servers decide to change unilaterally to $p'_{\tilde{X}_s}=(p'_{\tilde{X}_s}(k'))$, where $k'\in[k_1,k_2],\ p'_{\tilde{X}_s}(k')=1$, the clients who were bidding $k_2$ will follow the new price $k'$, since it leads to a higher payoff. Meanwhile, the clients who were bidding $k_1$ will choose a bid between $0$ and $k'$ by Fact \ref{fact3}. Bidding $k'$ yields a lower but positive payoff compared with the earlier, however, bidding $0$ yields a zero payoff. Thus, these clients will bid $k'$ as well.
The new trade ratio is then
\begin{equation*}
\begin{aligned}
\kappa' =& (p_{\tilde{X}_c}(k_1) + p_{\tilde{X}_c}(k_2)) p'_{\tilde{X}_s}(k')
=p_{\tilde{X}_c}(k_1)+p_{\tilde{X}_c}(k_2)>\kappa.
\end{aligned}
\end{equation*}
The proof above implies within the clients' financial ability, merging two server's bids always increases the trade ratio, which induces the result of Lemma \ref{lem:single_price}.
Note that the proof also follows in the case that $p_{\tilde{X}_c}, p_{\tilde{X}_s}$ are p.d.fs by replacing the summations with integrals.

\textbf{Proof of Lemma \ref{lem:finite_pieces}}
By Lemma \ref{lem:inc_v}, we have $v^*_{\rho}$ is monotonically increasing in $b$, which induces the monotonicity of $v_{c\_win}$ and $v_{c\_lose}$. Therefore, $v_{c\_win}$ and $v_{c\_lose}$ are bounded increasing on the closed interval $[k-\frac{s}{1+\alpha},k-\frac{s-k}{\alpha}]$. Define $g(b)\triangleq v_{c\_win}(b)-v_{c\_lose}(b)$ for $b\in[k-\frac{s}{1+\alpha},k-\frac{s-k}{\alpha}]$. We have $g(b)$ is of bounded variation, i.e. $g(b)$ has finite total variation on $[k-\frac{s}{1+\alpha},k-\frac{s-k}{\alpha}]$. Thus, the number of zero crossings of $g(b)$ over the closed interval is finite. Therefore, $\theta_{c,\rho}(b)$ is piecewise constant on $[0,k-\frac{s-k}{\alpha}]$ with a finite number of constant intervals. Within each of the intervals, $\theta_{c,\rho}(b)$ is constant and either $0$ or $k$. At the boundaries of these intervals, where $v_{c\_win}(b)=v_{c\_lose}(b)$, we have $\theta_{c,\rho}(b)=\{0,k\}$.

\subsection{Proofs in Section \ref{sec:mfe}} \label{app_mfe}
\textbf{Proof of Lemma \ref{lem:mc_pc}}
From (\ref{markov_tran}), for Borel set $B$, we have $\mathbb{P}(b[t+1]\in B|b[t]=b)\geq(1-\beta)\Psi(B)>0$, which satisfies the Doeblin condition. Then the budget chain is ergodic. The rest of the first part proof follows the results in Chapter 12, Meyn and Tweedie \cite{MeyTwe09}. Since the regeneration happens independently of the budget transition between two regenerations, we could further derive the relationship between $\pi_{z}(B)$ and $\pi_{z}^{(\tau)}(B|b)$, where $\tau$ is the first regeneration time after $t=0$ and $b(0)=b$ so that $\pi_{z}^{(\tau)}(\cdot | \cdot)$ is the $\tau$-step transition function without any regenerations.
\begin{equation*}
\begin{aligned}
\pi_{z}(B)
&=\sum_{\tau=0}^{\infty}(1-\beta)\beta^{\tau}\int\pi_{z}^{(\tau)}(B|b)d\Psi(b) \\
&=\sum_{\tau=0}^{\infty}(1-\beta)\beta^{\tau}\mathbb{E}_{\Psi}\big(\pi_{z}^{(\tau)}(B|B_{init})\big),
\end{aligned}
\end{equation*}
where we use the short-hand $\mathbb{E}_{\Psi}\big(\pi_{z}^{(\tau)}(B|B_{init})\big)$ to mean $\int\pi_{z}^{(\tau)}(B|b)d\Psi(b)$.

Since $\pi_{z}(\cdot)$ is the invariant budget distribution through the transition kernel defined in (\ref{markov_tran}), $\pi_{z}^{(\tau)}(B|b)$ is basically the $\tau$ -step transitions starting at $b_0=b$ without regenerations. If $B$ is a Lebesgue null-set, we have $\Psi(B)=0$ and in each of the $\tau$ step $\pi_{z}^{(\tau)}(B|b)=0$, therefore, $\pi_{z}(B)=0$.

\textbf{Proof of Lemma \ref{lem:v_lip}}
For any given $z$, by Theorem \ref{thm:vi_converge} there is a unique $v^*_{z}(\cdot)$ which is the unique fixed point of the contraction mapping $T^j_{z}$ with Lipschitz constant $\lambda \in (0,1)$. Rewriting (\ref{bell_op}) in terms of $z$ and $k$, we have
\begin{equation*}
\begin{aligned}
&(T^j_{z}f)(b)\\
&= \beta f(b) + p_s(1-z)(k-c_{serve}+\beta \Delta f_s(b,k,c_{serve}))+ \\
& \max_{x_c\in\mathcal{A}_k(b)} \bigg( p_c x_c \frac{s-k+c_{lose}+\beta \Delta f_c(b,s,k,\alpha)}{k},0 \bigg ) - p_c c_{lose}\\
\end{aligned}
\end{equation*}
where\\
\tab $\mathcal{A}_k(b)=[0,b+s/(1+\alpha)]\cap \{0,k\},$ \\
\tab $\Delta f_s(b,k,c_{serve}) = f(b+k-c_{serve}) -f(b),$ and \\
\tab $\Delta f_c(b,s,k,\alpha) = f(b+s-k-\alpha(k-b)^+)-f(b).$ \\
Taking the derivative with respect to $z$ using the Envelope Theorem, we have $T^j_{z}$ is Lipschitz continuous in $z$ with constant $k-c_{serve}$.

Pick $z_1$ and $z_2$, we have $\frac{||T^j_{z_1}v^*_{z_2}-T^j_{z_2}v^*_{z_2}||_{\infty}}{|z_1-z_2|}\leq k-c_{serve}$. Since $v^*_{z_2}$ is the unique fixed point of the contraction mapping $T^j_{z_2}$, we have $T^j_{z_2}v^*_{z_2} = v^*_{z_2}$. Then we have $\frac{||T^j_{z_1}v^*_{z_2}-v^*_{z_2}||_{\infty}}{|z_1-z_2|}\leq k-c_{serve}$. Applying $T^j_{z_1}$ $n$ times, given the contraction parameter $\lambda$, we have $\frac{||T^{(n+1)j}_{z_1}v^*_{z_2}-T^{nj}_{z_1}v^*_{z_2}||_{\infty}}{|z_1-z_2|}\leq\lambda^{n}(k-c_{serve})$. Also, we have the unique fixed point of $T^j_{z_1}$ being $v^*_{z_1}$. Letting $n\rightarrow \infty$, completes the proof as follows:
\begin{equation*}
\begin{aligned}
\frac{||v^*_{z_1}-v^*_{z_2}||_{\infty}}{|z_1-z_2|}&\leq \sum^{\infty}_{n=0}\frac{||T^{(n+1)j}_{z_1}v^*_{z_2}-T^{nj}_{z_1}v^*_{z_2}||_{\infty}}{|z_1-z_2|} \\
&\leq\frac{k-c_{serve}}{1-\lambda} 
\end{aligned}
\end{equation*}

\textbf{Proof of Theorem \ref{thm:theta_con}}
Given $z$ and $k$, the optimal value function can be rewritten as
\begin{equation*}
\begin{aligned}
&v^*_{z}(b)\\
&= \beta v^*_{z}(b) + p_s(1-z)(k-c_{serve}+\beta \Delta v_s(b,k,c_{serve}))+ \\
& \max_{x_c\in\mathcal{A}_k(b)} \bigg( p_c\big(\bm{1}_{x_c=k}(s-k+\beta \Delta v_c(b,s,k,\alpha))- \bm{1}_{x_c=0}c_{lose} \big) \bigg ) \\
&= \beta v^*_{z}(b) + p_s(1-z)(k-c_{serve}+\beta \Delta v_s(b,k,c_{serve}))+ \\
& \max_{x_c\in\mathcal{A}_k(b)} \bigg( p_c x_c \frac{s-k+c_{lose}+\beta \Delta v_c(b,s,k,\alpha)}{k},0 \bigg ) - p_c c_{lose}\\
\end{aligned}
\end{equation*}
where\\
\tab $\mathcal{A}_k(b)=[0,b+s/(1+\alpha)]\cap \{0,k\},$ \\
\tab $\Delta v_s(b,k,c_{serve}) = v^*_{z}(b+k-c_{serve}) -v^*_{z}(b),$ and \\
\tab $\Delta v_c(b,s,k,\alpha) = v^*_{z}(b+s-k-\alpha(k-b)^+)-v^*_{z}(b).$ \\
Define the increasing piecewise linear convex function $h_{z}(y)$:$\mathbb{R}\mapsto \mathbb{R}$ given by \\
\begin{equation*}
\begin{aligned}
h_{z}(y) =& \phi(z) + \max_{x_c\in\mathcal{A}_k(b)} \big(x_c y)_+, \\
\end{aligned}
\end{equation*}
where $(\cdot)_+:=\max(\cdot, 0)$ and
\begin{align*}
\phi(z)&:=\beta v^*_{z}(b) + p_s(1-z)\Big(k-c_{serve}\\
&\quad +\beta \Delta v_s(b,k,c_{serve})\Big) - p_c c_{lose}.
\end{align*}
By Lemma \ref{lem:v_lip}, we have $v^*_{z}(\cdot)$ is continuous in $z$ for all $x_c\in \mathcal{A}_k(b)$. Thus we have $\phi(z)$ is continuous in $z$ for all $x_c\in \mathcal{A}_k(b)$. By Berge's Maximum Theorem we have the correspondence,\\
\begin{equation*}
\begin{aligned}
& \mathcal{F}(y):= \arg \max_{x_c\in\mathcal{A}_k(b)} (x_c y)_+\\
\end{aligned}
\end{equation*}
is upper hemicontinuous in $z$. Note that $\theta_{c,z}(b)$ is given by
\begin{align}
\theta_{c,z}(b)=\mathcal{F}\left(p_c \frac{s-k+c_{lose}+\beta\Delta v_c(b,s,k,\alpha)}{k}\right).
\end{align}
Given the Lipschitz continuity of $v^*_{z}(\cdot)$ in $z$, we conclude for every state $b$, $\theta_{c,z}(b)$ is upper hemicontinuous in $z$.

\textbf{Proof of Theorem \ref{thm:pi_con}}
By Lemma \ref{lem:mc_pc}, given $z$, we have the Markov process of the budgets has a unique stationary distribution $\pi_{z}$. Here, we will prove the continuity of $\pi_{z}$ in $z$. By the Portmanteau Theorem, we only need to show that for any uniform converging sequence $z_n\rightarrow z$ and any open set $B$, $\liminf_{n\rightarrow\infty}\pi_{z_n}(B)\geq\pi_{z}(B)$. Thus, by Fatou's Lemma, we have
\begin{equation*}
\begin{aligned}
\liminf_{n\rightarrow\infty}\pi_{z_n}(B) &= \liminf_{n\rightarrow\infty}\sum_{\tau=0}^{\infty}(1-\beta)\beta^{\tau}\mathbb{E}_{\Psi}\big(\pi_{z_n}^{(\tau)}(B|B_{init})\big)\\
&\geq\sum_{\tau=0}^{\infty}(1-\beta)\beta^{\tau}\mathbb{E}_{\Psi}\big(\liminf_{n\rightarrow\infty}\pi_{z_n}^{(\tau)}(B|B_{init})\big)
\end{aligned}
\end{equation*}
To complete the proof, we need to show that
\begin{equation*}
\liminf_{n\rightarrow\infty}\pi_{z_n}^{(\tau)}(B|b)\geq\pi_{z}^{(\tau)}(B|b). \text{ for every } b\in B_{init}
\end{equation*}
The proof of the above holds by mathematical induction and the Skorokhod representation theorem. Details of a similar proof can be found in the appendix of \cite{ManRam13}.

\textbf{Proof of Theorem \ref{thm:gamma_con}}
Define the single-point inverse (lower inverse) $\theta_{c,z}^{-1}(0)=\{b\geq 0:0\in\theta_{c,z}(b)\}$ and also the upper inverse $\tilde{\theta}_{c,z}^{-1}(0)=\{b\geq0:\theta_{c,z}(b)=\{0\}\}$. By Lemma \ref{lem:finite_pieces}, we have $\theta_{c,z}^{-1}(0)$ consists of a finite number of closed subintervals in $[0,k-\frac{s-k}{\alpha}]$, and $\tilde{\theta}_{c,z}^{-1}(0)$ a finite number of open intervals (in $\mathbb{R}_+$) the closure of which is exactly $\theta_{c,z}^{-1}(0)$, with the difference being only finitely many points. Since $\pi_{z}$ is absolutely continuous with respect to Lebesgue measure we have
\begin{align*}
\pi_{z}\left( \theta_{c,z}^{-1}(0)\right) = \pi_{z}\left( \tilde{\theta}_{c,z}^{-1}(0)\right),
\end{align*}
so that $\theta_{c,z}^{-1}(0)$ is a continuity set of $\pi_{z}$ for every $z$.

From the definition of $\theta_{c,z}(\cdot)$ we have
\begin{align*}
& \theta_{c,z}^{-1}(0)\\
& =\left\{ b: h(b)\geq 0, v^*_{z}(h(b))-v^*_{z}(b) \leq \frac{k-s-c_{lose}}{\beta}\right\}  \\
& \quad \cup \{ b: h(b) \leq 0 \}\\
& \tilde{\theta}_{c,z}^{-1}(0)\\
& =\left\{ b: h(b)\geq 0, v^*_{z}(h(b))-v^*_{z}(b) < \frac{k-s-c_{lose}}{\beta}\right\}  \\
& \quad \cup \{ b: h(b) \leq 0 \},
\end{align*}
where we have
\begin{align*}
h(b) & = b+s-k-\alpha (k-b)^+ \\
\text{and } \{b: h(b) \leq 0 \} & = \left[0, k-\frac{s}{1+\alpha} \right]
\end{align*}

We know that $v^*_z(\cdot)$ is Lipschitz continuous in $z$ so that for all $\epsilon>0$ we have
\begin{align*}
\|v_{z'}-v_{z} \|_\infty\leq   L \epsilon \; \forall z' \in [z-\epsilon,z+\epsilon] \cap [0,1]\\
\|v_{z'}-v_{z} \|_\infty<   L \epsilon \; \forall z' \in (z-\epsilon,z+\epsilon) \cap [0,1].
\end{align*}
This then implies that for all $\{b\geq k-\tfrac{s}{1+\alpha}: h(b) \geq 0\}$ we have
\begin{align*}
& v^*_{z}(h(b))-v^*_{z}(b) -2L \epsilon \leq v^*_{z'}(h(b))-v^*_{z'}(b) \\
& \leq v^*_{z}(h(b))-v^*_{z}(b) +2L \epsilon \; \forall z' \in [z-\epsilon,z+\epsilon] \cap [0,1]\\
& v^*_{z}(h(b))-v^*_{z}(b) -2L \epsilon  < v^*_{z'}(h(b))-v^*_{z'}(b) \\
& < v^*_{z}(h(b))-v^*_{z}(b) +2L \epsilon \; \forall z' \in (z-\epsilon,z+\epsilon) \cap [0,1]
\end{align*}
Therefore we have
\begin{align*}
\theta_{c,z'}^{-1}(0) \subseteq F_{z}(\epsilon), \;
O_{z}(\epsilon) \subseteq \tilde{\theta}_{c,z'}^{-1}(0),
\end{align*}
where the closed set $F_{z}(\epsilon)$ and the open set $O_{z}(\epsilon)$ are given by
\begin{align*}
& F_{z}(\epsilon)\\
& =\left\{ b: h(b)\geq 0, v^*_{z}(h(b))-v^*_{z}(b) \leq \frac{k-s-c_{lose}}{\beta}+ 2L \epsilon \right\}  \\
& \quad \cup \{ b: h(b) \leq 0 \},\\
& O_{z}(\epsilon)\\
& =\left\{ b: h(b)\geq 0, v^*_{z}(h(b))-v^*_{z}(b) < \frac{k-s-c_{lose}}{\beta} - 2L \epsilon \right\}  \\
& \quad \cup \{ b: h(b) \leq 0 \}.
\end{align*}
Given a sequence $\{z_n\}_{n\geq 1}$ such that $\lim_{n\rightarrow\infty} z_n = z$, we know that $\pi_{z_n}\Rightarrow\pi_{z}$. Next fix an $\epsilon>0$. Then by the Portmanteau theorem we have
\begin{align*}
\pi_{z}(O_{z}(\epsilon)) & \leq \liminf_{n\rightarrow\infty} \pi_{z_n}(O_{z}(\epsilon))\\
& \leq \liminf_{n\rightarrow\infty} \pi_{z_n}(\tilde{\theta}_{c,z_n}^{-1}(0)) = \liminf_{n\rightarrow\infty} \pi_{z_n}(\theta_{c,z_n}^{-1}(0)) \\
& \leq \limsup_{n\rightarrow\infty} \pi_{z_n}(\tilde{\theta}_{c,z_n}^{-1}(0)) = \limsup_{n\rightarrow\infty} \pi_{z_n}(\theta_{c,z_n}^{-1}(0)) \\
& \leq \limsup_{n\rightarrow\infty} \pi_{z_n}(F_{z}(\epsilon)) \\
& \leq \pi_{z}(F_{z}(\epsilon)).
\end{align*}
Now the proof that $\lim_{n\rightarrow\infty} \pi_{z_n}(\theta_{c,z_n}^{-1}(0))=\pi_{z}(\theta_{c,z}^{-1}(0))$ follows by noticing that for $\epsilon_1 < \epsilon_2$ we have
\begin{align*}
O_{z}(\epsilon_2) &\subseteq O_{z}(\epsilon_1),\;
F_{z}(\epsilon_1)  \subseteq F_{z}(\epsilon_2), \\
\text{so }\lim_{\epsilon\downarrow 0} O_{z}(\epsilon) & = \tilde{\theta}_{c,z}^{-1}(0),\;
\lim_{\epsilon\downarrow 0} F_{z}(\epsilon) = \theta_{c,z}^{-1}(0),
\end{align*}
and also the fact that $\pi_{z}(\tilde{\theta}_{c,z}^{-1}(0)) = \pi_{z}(\theta_{c,z}^{-1}(0))$.



\end{document}